\documentclass[twocolumn]{aastex631}

\begin{document}

\title{Probing the Physical and Chemical Characteristics of an Extremely Early Class 0 Protostar in G204.4-11.3A2-NE 
}

\author{Hsuan-I (Ethan) Chou}
\altaffiliation{Current Address: Columbia University, 116th Street and Broadway, New York, NY 10027, USA; \url{hc3557@columbia.edu}}
\affiliation{Kang Chiao International School, No. 800, Huacheng Rd, New Taipei 231049, Taiwan (R.O.C.)}
\affiliation{Academia Sinica Institute of Astronomy and Astrophysics, 11F of Astronomy-Mathematics Building, AS/NTU, No.1, Sec. 4, Roosevelt Rd, Taipei 10617, Taiwan (R.O.C.)}

\author{Naomi Hirano}
\affiliation{Academia Sinica Institute of Astronomy and Astrophysics, 11F of Astronomy-Mathematics Building, AS/NTU, No.1, Sec. 4, Roosevelt Rd, Taipei 10617, Taiwan (R.O.C.)}

\author{Masayuki Yamaguchi}
\affiliation{National Astronomical Observatory of Japan, National Institutes of Natural Sciences, 2-21-1 Osawa, Mitaka, Tokyo 181-8588, Japan}
\affiliation{Department of Earth and Planetary Sciences, Faculty of Sciences, Kyushu University, Nishi-ku, Fukuoka 819-0395, Japan}
\affiliation{Academia Sinica Institute of Astronomy and Astrophysics, 11F of Astronomy-Mathematics Building, AS/NTU, No.1, Sec. 4, Roosevelt Rd, Taipei 10617, Taiwan (R.O.C.)}

\begin{abstract}
We have observed the low-mass molecular cloud core G204.4-11.3A2-NE (G204NE) in the direction of Orion B giant molecular cloud with the Atacama Large Millimeter/submillimeter Array in Band 6. 
The $\rm 1.3\ mm$ continuum images and visibilities unveil a compact central structure with a radius of $\sim$12 au, while showing no signature of binarity down to 18 au.
The bolometric temperature and luminosity of this source are derived to be ${\sim}$33 K and $\sim$1.15 $L_{\odot}$, respectively.
Chemical stratification is observed in dense gas tracers, with C$^{18}$O emission peaking at the continuum position surrounded by the spatially extended emission of N$_2$D$^+$ and DCO$^+$.
This implies that the core is in a very early evolutionary stage in which $\rm CO$ depletion occurs in most regions except for a small area heated by the central source. 
The envelope kinematics indicates a rotating and infalling structure with a central protostar mass of { 0.08--0.1} $M_{\odot}$.
The protostar drives a collimated outflow traced by CO, SiO, SO, and H$_2$CO, with misaligned blueshifted and redshifted lobes exhibiting a pair of bow-like patterns.
High-velocity jets, extending up to 720 au, are detected in CO, SiO, and SO lines.
The jet launching region is likely { within twice } of the dust sublimation zone. 
The absence of a binary signature suggests the outflows and jets are driven by a single protostar, although a close binary cannot be ruled out. 
The observed deflection of the outflows and jet is likely due to turbulent accretion in a moderately magnetized core.

\end{abstract}


\section{Introduction} \label{sec:intro}
Investigating the initial conditions of protostellar collapse is crucial for comprehending the formation process of low-mass stars.
Young stellar objects (YSOs) in the earliest stage of protostellar evolution—the Class 0 phase—are suitable objects in studying the initial condition of star formation because their natal cores are considered to retain these primordial characteristics.
 On the other hand, the circumstellar structure evolves significantly in this stage: observational studies have suggested that circumstellar disks start to form in the Class 0 phase (e.g., \citealp{Tobin2012, Murillo2013, Aso2017, Aso2018}). 
Furthermore, protostars begin to elevate the temperatures in their surrounding environments, consequently modifying the chemical properties of their enclosing envelopes (e.g., \citealp{Aikawa2012, HiranoLiu2014, Harsono2015, Tychoniec2021}). 
Energetic jets and outflows generated by protostars interact with their surrounding environment. These interactions can create cavities within their natal cores, initiate shock waves, and generate turbulence (e.g., \citealp{Bachiller1996, Frank2014, Bally2016, Lee2020}).

G204.4-11.3{ A2} located in the direction of Orion B giant molecular cloud is a cold gas clump ($T_{\rm dust}$$\sim$11 K) discovered by the Planck all-sky survey \citep{Planck2016}.
As shown in Figure \ref{fig:SED}{ c}, the James Clerk Maxwell Telescope (JCMT) Submillimeter Common-User Bolometer Array 2 (SCUBA-2) 850 $\mu$m image shows two peaks { connected by low-level emission stretching from the east to the southwest}.
The SW peak (G204SW) is associated with a bright source in the \emph{Wide-field Infrared Survey Explorer} (\emph{WISE}) 22 $\mu$m image, while the NE peak (G204NE), which is brighter in the 850 $\mu$m image, lacks a corresponding source at 22 $\mu$m.
The absence of a 22 $\mu$m source suggests that the G204NE molecular cloud core is at an early evolutionary stage.
Chemical properties of the core also support this hypothesis.
The N$_2$H$^+$ 1--0 emission peaks toward G204NE \citep{Tatematsu2017}.
In addition, G204NE shows a high abundance of deuterated species; the column density ratio of $N(\rm N_2\rm D^+)$/$N(\rm N_2\rm H^+)$ was measured to be 0.27 \citep{Kim2020}, which is comparable to those of the evolved prestellar cores \citep{Crapsi2005}.
{ Out of the 74 protostellar cores located within the Orion region ($\lambda$ Orionis, Orion A, and Orion B) studied by \citet{Kim2020}, only five cores, including G204NE, display a $N(\rm N_2\rm D^+)$/$N(\rm N_2\rm H^+)$ ratio exceeding 0.25.}
The observed physical properties (centrally peaked continuum emission and high column density) and chemical properties (N$_2$H$^+$ peak and high D/H ratio) suggest that G204NE is an ideal object for studying the initial condition of low-mass star formation.
In this paper, we report the Band 6 Atacama Large Millimeter/Submillimeter Array (ALMA) observations of G204NE. We present our investigation of the 1.3\ mm continuum and the molecular lines: $\rm N_2D^+\ 3–2$, $\rm CO\ 2–1$, $\rm SO\ 6_5–5_4$, $\rm C^{18}O\ 2-1$, $\rm H_2CO\ 3_{2,2}-2_{2,1}$, $\rm H_2CO\ 3_{0,3}-2_{0,2}$, $\rm DCN\ 3-2$, $\rm SiO\ 5-4$, and $\rm DCO^+\ 3-2$. 

The typical distance to the Orion B GMC is estimated to be $\rm \sim 403 \pm 4 \ pc$ \citep{Kounkel2018}. However, Lynds 1622 (L1622), a dark cloud located near G204NE, stands out with a much closer estimated distance of $\rm \lesssim 360\ pc$ \citep{Tobin2020}. The radial velocity of L1622 was measured to be $v_{\rm{LSR}}\sim \rm 1~km~s^{-1}$ \citep{Maddalena1986}, which is different from that of Orion B, $v_{\rm{LSR}}\sim \rm 10~km~s^{-1}$. Since the systemic velocity of G204NE, $v_{\rm{LSR}}\sim \rm 1.4 ~km~s^{-1}$ (see Section \ref{subsec: kinematics}), is close to that of L1622 rather than that of Orion B, we assume that G204NE is also on the near side of the Orion B GMC and adopt {  360~pc} as the distance in this paper. 

The outline of this paper is as follows. Details regarding the ALMA $\rm 1.3\ mm$ observations, as well as the Submillimeter Array\footnote{The Submillimeter Array is a joint project between the Smithsonian Astrophysical Observatory and the Academia Sinica Institute of Astronomy and Astrophysics and is funded by the Smithsonian Institution and the Academia Sinica.} (SMA; \citealp{Ho2004}) $\rm 1.1\ mm$ and Atacama Compact Array (ACA) $\rm 830\ \mu m$ observations used for the spectral energy distribution (SED) analysis (see Section \ref{subsec: SED}), are summarized in Section \ref{sec:obs}. In Section \ref{sec:results}, we present the results of the SED, the continuum, and the molecular lines. Analyses of molecular abundances, envelope kinematics, and the jets and outflows are presented in Section \ref{sec:analysis}. The implications of our results and analyses are discussed in Section \ref{sec:discussion}. Finally, we summarize the conclusions of this study in Section \ref{sec:conclusions}.

\section{Observations} \label{sec:obs}
\subsection{ALMA 1.3 mm} \label{subsec:ALMA}
\begin{deluxetable*}{cccccc} \label{tab:lines}
\tablecaption{Summary of ALMA observational parameters.}
\tablewidth{\textwidth}
\tablehead{\colhead{Line/continuum} & \colhead{Rest frequency} & \colhead{Spectral resolution$^{\rm a}$} & \colhead{rms noise level} & \colhead{Beam size (P.A.)} \\ 
\colhead{} & \colhead{(GHz)} & \colhead{($\rm km\ s^{-1}$)} & \colhead{($\rm mJy\ beam^{-1}$)} & \colhead{} } 
\startdata
TM1 Continuum (CLEAN) & $232.996$ & - & $0.024$ & $\rm 0''.087 \times 0''.070$ ($-86^{\circ}$)\\
TM1 Continuum (SpM) & $232.996$ & - & - & $0''.055 \times 0''.046$ ($83^{\circ}$)$\rm ^b$ \\
TM1+TM2 Continuum (CLEAN) & $232.996$ & - & $0.061$ & $\rm 0''.32 \times 0''.28$ ($-90^{\circ}$)\\
$\rm N_2D^+\ 3-2$ & $231.322$ & $\rm 0.158$ & $3.9^{\rm c}$ & $\rm 0''.31 \times0''.28$ ($-83^{\circ}$)  \\ 
$\rm CO\ 2-1$ & $230.538$ & $\rm 0.184$ & $4.0^{\rm d}$ & $\rm 0''.31\times0''.28$ ($-84^{\circ}$) \\ 
$\rm SO\ 6_5-5_4$   & $219.949$ & $\rm 0.166$ & $3.0^{\rm d}$ & $\rm 0''.31\times0''.28$ ($-84^{\circ}$) \\ 
$\rm C^{18}O\ 2-1$ & $219.560$ & $0.167$ & $2.5^{\rm c}$ & $\rm 0''.31\times0''.28$ ($-81^{\circ}$)   \\ 
$\rm H_2CO\ 3_{22}-2_{21}$ & $218.476$ & $\rm 0.168$ &  $2.1^{\rm d}$ &  $\rm 0''.31\times0''.28$ ($-83^{\circ}$) \\ 
$\rm H_2CO\ 3_{03}-2_{02}$ & $218.222$ & $\rm 0.168$ &  $2.1^{\rm d}$ &  $\rm 0''.31\times0''.28$ ($-78^{\circ}$) \\ 
$\rm DCN\ 3-2$ & $217.239$ & $\rm 0.168$ & $\rm 1.4^{\rm c}$ & $\rm 0''.31\times0''.28$ ($-83^{\circ}$)  \\ 
TM1 $\rm SiO\ 5-4$ & $217.105$ & $\rm 0.195$ & $1.4^{\rm d}$ &  $\rm 0''.11\times0''.083$ ($83^{\circ}$) \\ 
TM1+TM2 $\rm SiO\ 5-4$ & $217.105$ & $\rm 0.195$ & $2.0^{\rm { d}}$ &  $\rm 0''.31\times0''.28$ ($-84^{\circ}$) \\ 
$\rm DCO^{+}\ 3-2$   & $216.113$ & $\rm 0.169$ & $2.4^{\rm c}$ & $\rm 0''.31\times0''.28$ ($-80^{\circ}$) \\ 
\enddata
\tablenotetext{a}{The spectral resolution is a factor of two larger than the channel spacing.}
\tablenotetext{b}{The effective ``beam size'' of the SpM image was estimated using the ``point-source injection'' method (see Section \ref{subsec:ALMA}).}
\tablenotetext{c}{Measured in the data cube with a channel width of 0.1 km s$^{-1}$. }
\tablenotetext{d}{Measured in the data cube with a channel width of 0.2 km s$^{-1}$.}
\end{deluxetable*}
We observed G204NE with ALMA in Band 6 during Cycle 4 (Project ID: 2017.1.00707.S, PI: Naomi Hirano). Two different array configurations were used: 12 m C43-5 (TM1) and 12 m C43-2 (TM2). The four execution blocks (EBs) of the TM1 observations were carried out in 2017 on November 26, December 3, and December 6 with 50, 47, and 45 antennas, respectively. The two EBs of the TM2 were carried out in 2018 on April 11 and 18, with 43 and 45 antennas, respectively. The field center of the observations is $\rm \alpha(J2000) = 05^{h}55^{m}38^{s}.430$, $\rm \delta(J2000) = 2^{\circ}11'33''.30$. The projected baseline lengths of the two array configurations range from $\rm 15.1\ m$ to $\rm 8547.6\ m$ for TM1 and from $\rm 15.1\ m$ to $\rm 500.2\ m$ for TM2. The maximum recoverable scale of the TM1 dataset was $1''.6$, and that of the TM2 dataset was $8''.1$. The correlator was configured to accommodate nine molecular lines: $\rm N_2D^+\ 3-2$, $\rm CO\ 2-1$, $\rm SO\ 6_5-5_4$, $\rm C^{18}O\ 2-1$, $\rm H_2CO\ 3_{2,2}-2_{2,1}$, $\rm H_2CO\ 3_{0,3}-2_{0,2}$, $\rm DCN\ 3-2$, $\rm SiO\ 5-4$, and $\rm DCO^+\ 3-2$. The spectral resolutions of the lines were either $122~\rm kHz$ or $141~\rm kHz$, which correspond to velocity resolutions of $\sim 0.16-0.2~\rm km~s^{-1}$ as listed in Table \ref{tab:lines}. The total bandwidths were either $58.6~\rm MHz$ ($\sim 80\rm ~km~s^{-1}$) or $117.2~\rm MHz$ ($\sim 150-160\rm ~km~s^{-1}$). In addition, a broad-band spectral window with a bandwidth of $1.875~\rm GHz$ and a spectral resolution of $31.25~\rm MHz$ centered at a frequency of $233.0~\rm GHz$ was dedicated to observing the continuum emission.

\subsubsection{Calibration and CLEAN Imaging}
Both TM1 and TM2 data were calibrated using the Common Astronomical Software Applications (CASA; \citealp{CASA2022}) version 5.1.1 and the ALMA Pipeline 40896 (Pipeline-CASA51-P2-B). The TM1 observations used $\rm J0510-1800$ as bandpass and flux calibrators, while the TM2 observations used $\rm J0522-3627$ and $\rm J0423-0120$ as both bandpass and flux calibrators. For both TM1 and TM2, $\rm J0552+0313$ was used as the gain calibrator. 

In addition, we performed self-calibration and imaging on the continuum data using the tasks $\tt tclean$, $\tt gaincal$, and $\tt applycal$ in CASA version 6.5.0. 
We applied self-calibration onto two visibility datasets: one with only TM1 and another with TM1 and TM2 combined (hereafter ``TM1+TM2''). 
{ For TM1 data, three rounds of self-calibration were performed: the first and second steps are phase calibrations with solution intervals of $\tt inf$, and the third step is amplitude and phase calibrations with solution intervals of $\tt inf$. For TM1+TM2 data, { two} rounds of phase self-calibration were performed with solution intervals of $\tt inf$. Self-calibration improved the signal-to-noise ratios ($\rm S/N$) of both TM1 and TM1+TM2 continuum maps by $\sim 30\%$ compared to those of the non-self-calibrated CLEAN images. }
The obtained calibration table was also applied to the line visibility data.

Briggs weighting with a robust parameter of $+0.5$ was applied to both datasets in $\tt tclean$, and CLEAN masks were determined by the auto-masking algorithm. 
The parameters employed for the auto-masking procedure were the standard values\footnote{{ \url{https://casaguides.nrao.edu/index.php/Automasking\_Guide\_CASA\_6.5.4}}} pertinent to the 12 m array with the 75th percentile of baselines exceeding 400 m.
Due to the non-uniform uv-coverage provided by the two different array configurations, the dirty beam of the TM1+TM2 image exhibits a non-Gaussian shape with a substantial shelf surrounding the central Gaussian core. 
{We tried to compare the beam shapes of the TM1+TM2 continuum images using different uv taper values, and found that the one with a uv taper of $0''.2$ could provide reasonably good beam shape without degrading the angular resolution.}
The synthesized beam size of the TM1 continuum map is {$\sim 0''.078$}, which corresponds to $\sim 28~\rm au$ at a distance of $360~\rm pc$. 
The beam size of the TM1+TM2 continuum map { with a uv taper of $0''.2$} is { $\sim 0''.30$} ($\sim 108~\rm au$). 
Both of the continuum maps were primary beam corrected with $\tt impbcor$. 

We obtained the line data cubes from the TM1+TM2 visibility data in CASA version 6.1.0. The continuum was subtracted from the line data using the task $\tt uvcontsub$. 
{ Line-free channels were determined in the time-averaged spectra of the TM2 visibility data.}
Using the $\tt tclean$ task, we applied Briggs weighting with a robust parameter of $\rm +0.5$, as well as a UV taper of $0''.2$. The synthesized beam sizes of the TM1+TM2 line data cubes are { $\sim 0''.29$}, which corresponds $\rm \sim 108 \ au$. In order to spatially resolve the compact jet components (see Section \ref{subsubsec:outflow_tracers}), we made a higher-resolution SiO data cube using the TM1 data without UV taper. The synthesized beam of the high-resolution $\rm SiO$ data cube was { $\sim 0''.096$} ($\sim 36~\rm au$). The $\rm C^{18}O$, $\rm N_2D^+$, and $\rm DCO^+$ lines were imaged every $\rm 0.1\ km\ s^{-1}$, while the $\rm H_2CO$, $\rm SO$, $\rm SiO$, and $\rm CO$ lines were imaged every $\rm 0.2\ km\ s^{-1}$. All line images presented in this paper have been primary beam corrected using the task $\tt impbcor$. 

{The synthesized beam size and the rms noise level per channel of each molecular line are summarized in Table \ref{tab:lines}. }
The rms noise levels were measured before primary beam correction.

\subsubsection{Super-resolution Imaging with Sparse Modeling}

We used $\tt PRIISM$ \footnote{$\tt PRIISM$ (Python Module for Radio Interferometry Imaging with Sparse Modeling) is an imaging tool for ALMA based on the sparse modeling technique publicly available at \url{https://github.com/tnakazato/priism}.} version 0.11.5 \citep{Nakazato2020} to image a higher-resolution TM1 continuum map with sparse modeling (SpM). This package involves regularized maximum-likelihood optimizations with $\ell _1$+TSV imaging and the cross-validation (CV) scheme as illustrated in \citet{Yamaguchi2020, Yamaguchi2021}. The model image is reproduced by minimizing a cost function, which is the chi-square error between the visibility model derived by the model image and observed visibility. This is accompanied by two regularization terms, $\ell_1$-norm and the total squared variation (TSV), which are parameterized by $\Lambda_{l}$ and $\Lambda_{tsv}$, respectively. With the self-calibrated TM1 visibility data, we obtain a model image that minimizes the cost function for a given set of $(\Lambda_{l}, \Lambda_{tsv})$ and has a set of model images with a range of values of $(\Lambda_{l}, \Lambda_{tsv})$. To select a pair of $(\Lambda_{l}, \Lambda_{tsv})$ for the optimal image, we employ the 10-fold cross-validation (CV) approach. The model with minimum cross-validation error (CVE) is set to the optimal one \citep{Yamaguchi2020}. The final pairs for $(\Lambda_{l}, \Lambda_{tsv})$ with the minimum CVE provided ($10^{5}$, $3\times10^{13}$). The final model image has units of $\rm Jy\ pixel^{-1}$ instead of $ \rm Jy\ beam^{-1}$ because this imaging process does not involve beam convolution. 

We evaluated the effective spatial resolution $\theta_{\rm eff}$ of the SpM image using the ``point-source injection'' method outlined in \cite{Yamaguchi2021}. We injected an artificial point source into the observed visibility data. The flux density of the point source was set to $10\%$ of the total flux density of the target. The artificial point source was north-placed in an emission-free area but at a distance within the maximum recoverable scale ($< 0''.7$ in radius). The imaging was performed for the point-source injected data using the same set of regularization parameters employed for generating the optimal image of non-injected data. The intensity distribution of each injected point source was Gaussian-like in the reconstructed image; therefore, we fitted it with an elliptical Gaussian function to measure the full widths at half maxima (FWHMs), which can be regarded as an effective spatial resolution. The effective spatial resolution is $0''.055\times 0''.046$ ($\rm \sim 18~au$; $\rm P.A. = 83^{\circ}$), which is an improvement over that of the TM1 CLEAN image by a factor of $\sim 1.6$. The information regarding the SpM image above is summarized in Table \ref{tab:lines}.

We were unable to measure the noise level of the SpM continuum map, as the background noise pattern of the image does not follow the Gaussian distribution.

\subsection{SMA 1.1 mm and 1.3 mm} \label{subsec:SMA}
The $\rm 1.1~mm$ ($\rm 280~GHz$) and $\rm 1.3~mm$ ($\rm 230~GHz$) continuum observations were carried out in 2017 on April 21 with the compact configuration of the SMA (Project ID: 2016B-A013; PI: Naomi Hirano). The two frequency bands were observed simultaneously using the $\rm 240~GHz$ and $\rm 345~GHz$ receivers. The frequency coverages of the $\rm 240~GHz$ receiver were from $215.5$ to $\rm 221.5~GHz$ in the lower sideband (LSB) and from $\rm 229.5$ to $\rm 235.5~GHz$ in the upper sideband (USB), while those of the $\rm 345~GHz$ receiver were from $\rm 264.5$ to $\rm 270.5~GHz$ in the LSB and from $\rm 278.5$ to $\rm 284.5~GHz$ in the USB. The field center of the observations was $\rm \alpha(J2000) = 05^h55^m38^s.400$, $\rm \delta(J2000) = 02^{\circ}11'35''.50$. 

Calibration was performed using the MIR/IDL software package \footnote{\url{https://www.cfa.harvard.edu/rtdc/SMAdata/process/mir/}}. $\rm J0510+180$ and $\rm J0607-085$ were used as the gain calibrators, $\rm 3C\ 279$ was used as the bandpass calibrator, and Callisto was used as the flux calibrator. 
The continuum image was obtained in CASA version 6.1.0 { using the $\tt tclean$ task with} Briggs weighting and a robust parameter of $+0.5$. 
The continuum emission was obtained by averaging the line-free channels. To improve the $\rm S/N$, we combined the USB and LSB data. The SMA synthesized beams are  $\rm 2''.8 \times 2''.3$ ($\rm P.A. = 72^{\circ}$) and $\rm 3''.4 \times 2''.9$ ($\rm P.A. = -88^{\circ}$) for the $\rm 280~GHz$ and $\rm 230~GHz$ images, respectively. The rms noise levels were measured by including emission-free regions to be $\rm \sim 2.2~mJy~beam^{-1}$ and $\rm \sim 2.4~mJy~beam^{-1}$ for the $\rm 280~GHz$ and $\rm 230~GHz$ images, respectively.

\subsection{ACA 830 $\mu m$} \label{subsec:ACA}
The ACA Band 7 observations of G204NE were carried out at $\rm 830\ \mu m$ during Cycle 7 (Project ID: 2021.1.00727.S; PI: Naomi Hirano). The three EBs in 2021 on November 8, 11, and 29 used 10, 9, and 8 antennas of $\rm 7\ m$ diameters, respectively. The field center of the observations is $\rm \alpha(J2000) = 05^h34^m55^s.840$, $\rm \delta(J2000) = -05^{\circ}46'05''.00$. The projected baseline lengths range from $\rm 8.9\ m$ to $\rm 45\ m$. The observations used a correlator configuration for the continuum centered at $\rm 357.959\ GHz$ with a bandwidth of $\rm 2\ GHz$, as well as the molecular lines $\rm DCO^+\ 5-4$ ($\rm 360.138\ GHz$), $\rm H_2D^+\ 1_{1,0}-1_{1,1}$ ($\rm 372.388\ GHz$), and $\rm N_2H^+\ 4-3$ ($\rm 372.639\ GHz$) with a bandwidth of $\rm 0.25\ GHz$ and $4096$ channels. However, we defer the discussion of these molecular lines to future papers. 

Calibration was performed in CASA version 6.2.1-7 (Pipeline 2021.2.0.128). J0541-0541 was used as the gain calibrator, while J0423-0120 was used as the bandpass and flux calibrators. 
{After checking the weblog, we decided to use the clean images produced by the pipeline. 
The pipeline adopted the Briggs weighting with a { robust} parameter of $+$0.5.
The ACA synthesized beam is $4''.9 \times 3''.0$ ($\rm P.A.=-71.2^{\circ}$).
The rms noise level measured from the emission-free region is 10 mJy beam$^{-1}$. 
The maximum recoverble scale of the ACA image is 19\farcs7.}

\begin{deluxetable}{ccc} \label{tab:SED_measurements}
\tablewidth{\textwidth}
\tablecaption{Flux measurements given at different wavelengths for SED.}
\tablehead{\colhead{Observation} & \colhead{Total flux density} & \colhead{Uncertainty}\\ 
\colhead{ } & \colhead{($\rm mJy$)} & \colhead{($\rm mJy$)}}
\startdata
\emph{WISE} $\rm 3.4~\mu m$ & $<0.012$ & {-}\\
\emph{WISE} $\rm 4.6~\mu m$ & $0.374$ & {$0.017$}\\
\emph{WISE} $\rm 12~\mu m$ & $<0.746$ & {-}\\
\emph{WISE} $\rm 22~\mu m$ & $<2.47$ & {-}\\
\emph{Herschel} $\rm 70~\mu m$ & $1490$ & {$30$}\\
\emph{Herschel} $\rm 160~\mu m$ & $9730$ & {$170$}\\
\emph{Herschel} $\rm 250~\mu m^{a}$ & $10400$ & {$400$}\\
\emph{Herschel} $\rm 350~\mu m^{a}$ & $9160$ & {$240$}\\
\emph{Herschel} $\rm 500~\mu m^{a}$ & $6390$ & {$230$}\\
ACA $\rm 830~\mu m$ & ${ 666}^{\rm b}$ & {66$^{\rm c}$}\\
JCMT $\rm 850~\mu m$ & ${738}^{\rm b}$ & {37$^{\rm d}$}  \\
SMA $\rm 1.1~mm$ & ${200}^{\rm e}$ & {20$^{\rm c}$}\\
SMA $\rm 1.3~mm$ & ${139}^{\rm e}$ & {14$^{\rm c}$}\\
ALMA $\rm 1.3~mm$ & ${157}^{\rm f}$ & {16$^{\rm c}$} \\
\enddata
\tablenotetext{a}{Not included in the SED.}
\tablenotetext{b}{The flux was measured using a circular aperture with a radius of $\sim 10''$.}
\tablenotetext{c}{The absolute flux error of 10\%.}
\tablenotetext{d}{The absolute flux error of 5\%.}
\tablenotetext{e}{The flux was measured using a circular aperture of $5''$.}
\tablenotetext{f}{The flux was measured using a circular aperture of $5''$ on the TM2 continuum image.}
\end{deluxetable}

\section{Results} \label{sec:results}
\subsection{Spectral Energy Distribution} \label{subsec: SED}
\begin{figure*}[ht!]
\includegraphics[width=\textwidth]{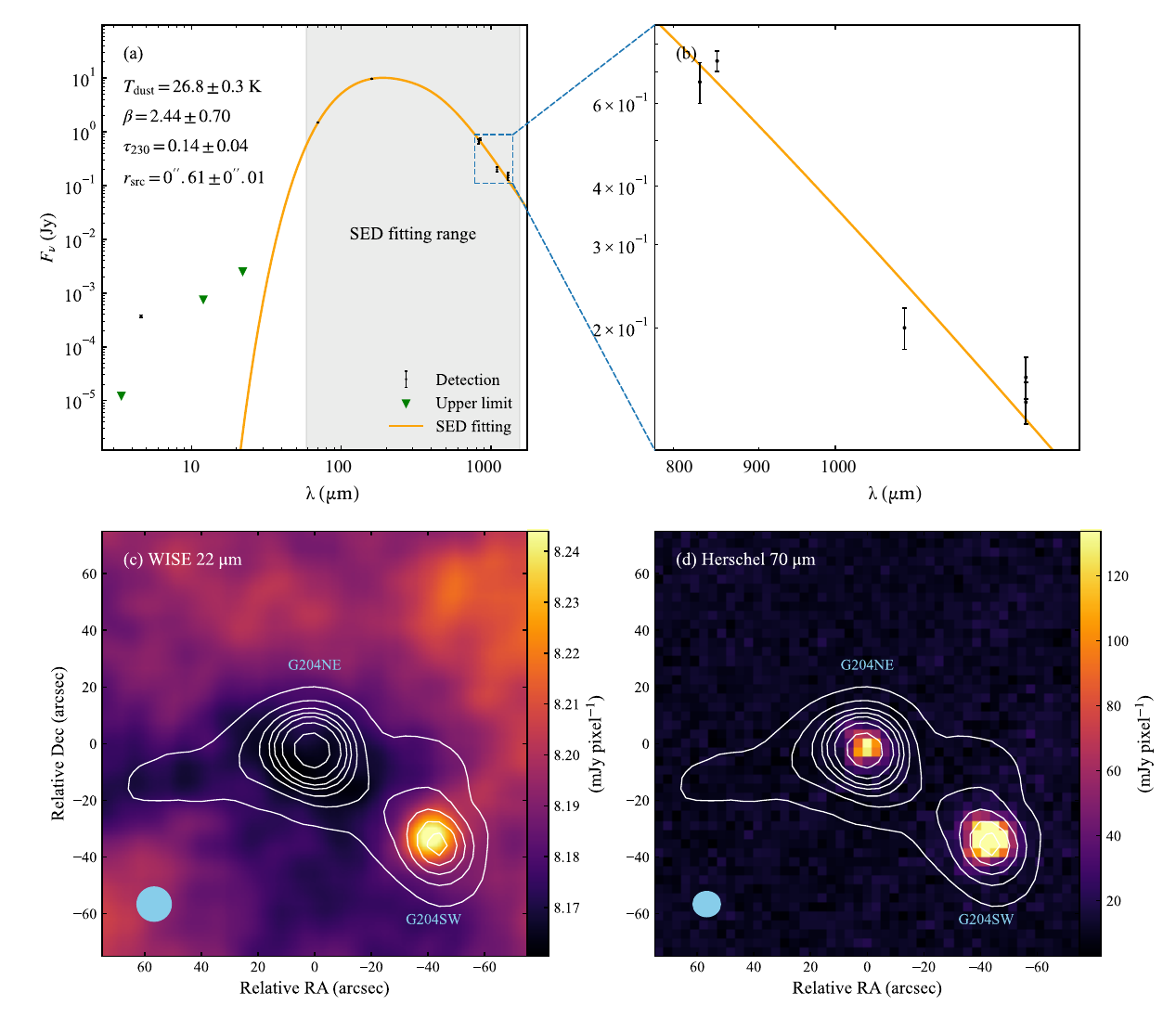}
\caption{{(a) Spectral energy distribution plot of G204NE, with detections (black circular points), $3 \sigma$ upper limits (green triangle), and a modified blackbody fit (orange curve). The gray shaded region indicates the range of measurements used for the fitting. (b) Zoom-in of the millimeter/submillimeter range. }(c) and (d) JCMT SCUBA-2 $\rm 850\ \mu m$ continuum image of the G204 region (contour) overlaid on the (c) \emph{WISE} $\rm 22\ \mu m$ and (d) \emph{Herschel} $\rm 70\ \mu m$ images (color). 
The contour levels  are from $\rm 3\sigma$ to $\rm 15\sigma$ in steps of $\rm 3\sigma$ and then from $\rm 20\sigma$ to $\rm 40\sigma$ in steps of $\rm 10\sigma$, where $\sigma$ corresponds to $\rm 17.3\ mJy\ beam^{-1}$.  
The blue-filled ellipses in (c) and (d) show the point-spread functions:  $\rm 12'' \times 12''$ for \emph{WISE} $\rm 22\ \mu m$ and $\rm 8''.8 \times 9''.6$ for \emph{Herschel} $\rm 70\ \mu m$, respectively. 
\label{fig:SED}}
\end{figure*}

Figure \ref{fig:SED}(a) shows the {spectral energy distribution (SED)} plot derived from the \emph{WISE} $\rm 3.4$, $\rm 4.6$, $\rm 12$, and $\rm 22$ $\rm \mu m$; \emph{Herschel} $\rm 70$ and $\rm 160~\mu m$; ACA $\rm 830~\mu m$; JCMT SCUBA-2 $\rm 850~\mu m$ \citep{Tatematsu2017}; SMA $\rm 1.1$ and $\rm 1.3~mm$; and ALMA $\rm 1.3~mm$ observations. 
The flux measurements of the \emph{WISE} and \emph{Herschel} data were obtained from the SIMBAD database (url: \url{simbad.u-strasbg.fr/}).
{The flux measurements of the ACA, JCMT, SMA, and ALMA were performed using the circular aperture. 
We quote a nominal 10\% absolute flux error for the interferometric data and a 5\% for SCUBA-2 data, as they are larger than the 1${\sigma}$ thermal noise levels.}
We excluded the \emph{Herschel} $\rm 250$, $\rm 350$, and $\rm 500$ $\rm \mu m$ observations as the flux measurements at these wavelengths were contaminated by the extended emission from the cloud due to larger beam sizes. The flux measurements for the SED are listed in Table \ref{tab:SED_measurements}. 
{The 830 $\mu$m flux measured with the ACA is lower than the 850 $\mu$m flux measured with the JCMT.
This discrepancy arises from the inherent tendency of the interferometer to filter out spatially extended emission.
However, the difference is $\sim$10\%, which is not significant if the flux uncertainties are taken into account.
}

{The SED at the wavelength longer than 70 $\mu$m was fitted with the form of greybody radiation with a single emissivity, $\beta$, and a temperature, $T_{\rm dust}$,
\begin{equation}
 F_\nu = \Omega { B_\nu(T_{\rm dust}})(1-e^{-{\tau}_{\nu}})    \end{equation}
 where F$_{\nu}$ is the flux density, $\Omega$ is the solid angle of the emitting region, ${ B_\nu(T_{\rm dust})}$ is the { Planck} function, $T_{\rm dust}$ is the dust temperature, and $\tau_\nu$ is the optical depth of dust that is assumed to be proportional to $\nu^\beta$. 
The derived parameters, $T_{\rm dust}$, $\beta$, $\tau_{\rm 230 GHz}$, and the source { radius} $r = \sqrt{\ln {2}  \Omega / \pi}$ are 26.8$\pm$0.3 K, 2.44$\pm$0.70, 0.14$\pm$0.04, and 0\farcs61$\pm$0\farcs01, respectively.
It should be noted that the beta value derived from the current dataset lacks reliability, as it { suddenly} changes to  $\sim${ 1.2} when { an assumed} source size { exceeds} 0\farcs76.
To accurately derive the beta value, the data points at lower frequency bands are required.
On the other hand, the dust temperature is invariant with respect to the source size, remaining consistently between 25--27 K.}

As shown in Figures \ref{fig:SED}(c) and (d), G204NE was detected in the \emph{Herschel} $\rm 70\ \mu m$ observation but not in the \emph{WISE} $\rm 22\ \mu m$ observation. At $\rm 70\ \mu m$, G204NE is fainter than the neighboring core G204SW, which contains the $\rm 22\ \mu m$ source. On the other hand, G204NE is brighter than G204SW at $850\rm ~\mu m$. Although G204NE is generally not detected at wavelengths of $\lesssim \rm 22$ $\rm \mu m$, some faint emission was detected in the \emph{WISE} $\rm 4.6\ \mu m$ observation (Figure \ref{fig:SED}(a)).

Following \cite{Myers1993}, we used the flux-weighted mean frequency $\bar{\nu}=\int_0^\infty \nu F_\nu d\nu / \int_0^\infty F_\nu d\nu$ in the observed SED to estimate the bolometric temperature $T_{\rm bol}$, which can be given as:
\begin{equation}
    T_{\rm{bol}} = 1.25 \times 10^{-11} \bar{\nu}\ (\rm{K\ Hz^{-1}})
\end{equation}
Similarly, the bolometric luminosity can also be obtained through integration:
\begin{equation}
    L_{\rm{bol}} = 4\pi D^2 \int_{0}^{\infty} F_\nu d\nu
\end{equation}
where ${ D = 360\ \rm{pc}}$ is the distance to G204NE.
{We evaluated the integrals using trapezoidal Riemann sums over the flux measurements rather than fitting. To estimate the uncertainties in $T_{\rm bol}$ and $L_{\rm bol}$ due to the flux uncertainties, we employed a Monte Carlo approach. Each detected flux was modeled as a Gaussian distribution centered on the measured value, with a standard deviation equal to its uncertainty. For flux upper limits, we assumed a truncated Gaussian distribution centered at zero, with a standard deviation equal to one-third of the quoted upper limit (i.e., corresponding to a $3\sigma$ threshold), and truncated between zero and the limit value. We then generated $10^5$ synthetic realizations of the SED by randomly sampling from these distributions and measured the $T_{\rm bol}$ and $L_{\rm bol}$ for each realization. The resulting distributions of the measurements of $T_{\rm bol}$ and $L_{\rm bol}$ were found to be approximately Gaussian; hence, the means of the $T_{\rm bol}$ and $L_{\rm bol}$ measurements were adopted as the final values, and their standard deviations ($1\sigma$) were taken as the uncertainties. This provided $T_{\rm{bol}} = \rm{32.7\pm0.2~K}$ and $L_{\rm bol}= 1.15\pm0.02~L_{\odot}$, placing G204NE below the $T_{\rm bol} = \rm 70\ K$ threshold between the Class 0 and Class I phases \citep{Chen1995}. 
The flux measurements at 1.1 mm and 1.3 mm were acquired using interferometers that lack sensitivity to extended structures. 
Nonetheless, the impact of this filtering issue is unlikely to substantially affect the estimation of $T_{\rm bol}$ and $L_{\rm bol}$. 
The flux disparity between ACA (830 $\mu$m) and JCMT (850 $\mu$m) is only $\sim$10\%. 
The $T_{\rm bol}$ and $L_{\rm bol}$ values remain at 33.0 K and 1.15 $L_{\odot}$, when the flux measurements at 1.1 mm and 1.3 mm are doubled, which corresponds to the assumption that half of the flux was lost.
}

\subsection{1.3 mm Continuum Emission} 
\label{subsec:continuum}
\begin{figure*}[ht!]
\includegraphics[width=\textwidth]{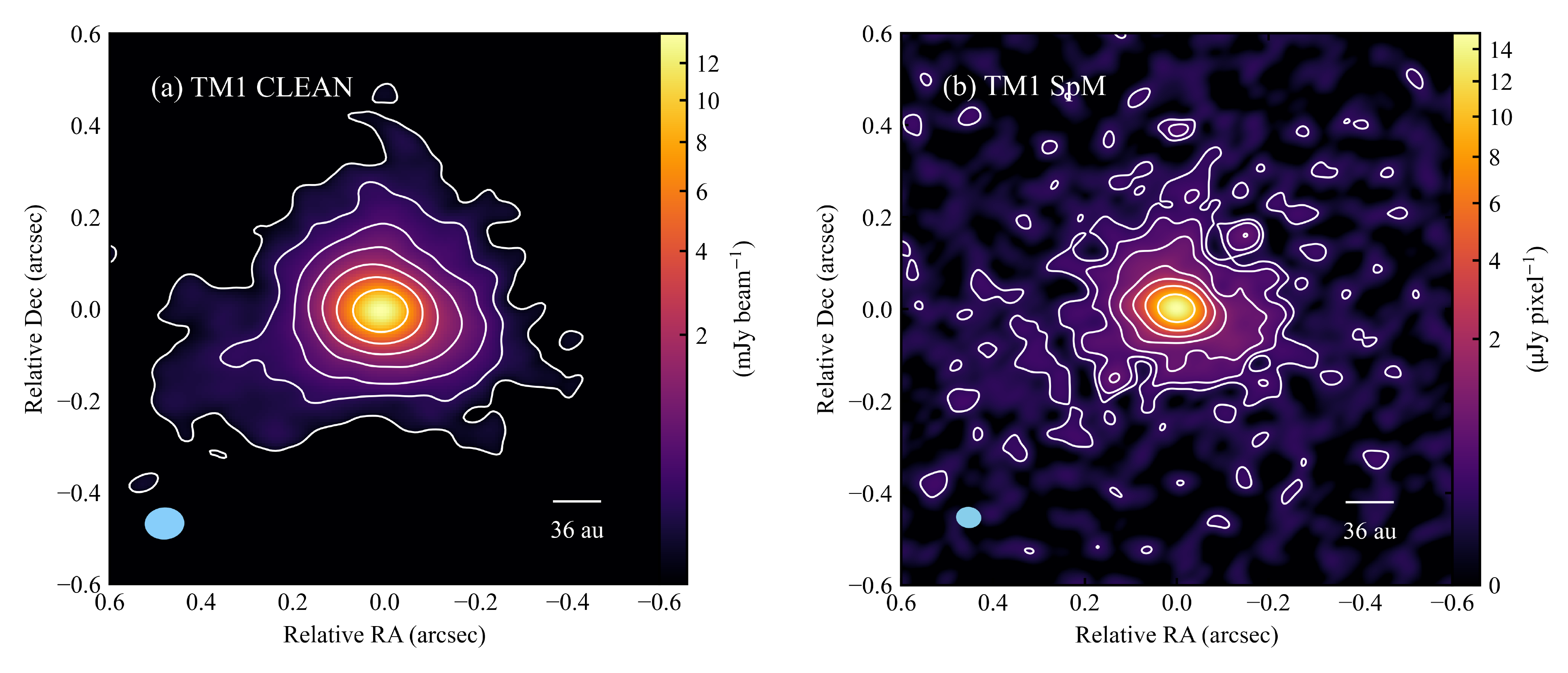}
\caption{1.3\ mm TM1 continuum emission maps imaged using (a) CLEAN and (b) SpM. The color scales on both images follow a power law with a scaling exponent of $\gamma=0.4$. The contour levels in (a) are 5, 10, 20, 40, … $\times \sigma$, where $\sigma$ corresponds to $\rm 0.024~mJy~beam^{-1}$. The contour levels in (b) are 0.01, 0.02, 0.04, … $\times$ the maximum intensity of $\rm 15.1~\mu Jy~pixel^{-1}$. The blue-filled ellipses in the bottom left corners of (a) and (b) denote the ALMA synthesized beam of the CLEAN image and the effective resolution of the SpM image, respectively. \label{fig:continuum}}
\end{figure*}

Figures \ref{fig:continuum}(a) and (b) show the TM1 $\rm 1.3\ mm$ continuum emission maps of G204NE generated with CLEAN and SpM, respectively, which have higher resolutions more suitable for inspecting the compact emissions of the source than the TM1+TM2. In the CLEAN image, the emission exhibits a structure with a radius of $\rm \sim 0''.2$ ($\rm 72~au$) at the $10\sigma$ contour level. 
The {intensity} at the peak is $\rm \sim 14\ mJy\ beam^{-1}$.  
This corresponds to a brightness temperature of $\sim 52~\rm K$, which can be regarded as the dust temperature by assuming optical thickness. No substructures or close binaries are detected, even in the SpM image with an effective resolution of $\sim 0''.05$, although we do not dismiss the possibility of such components being spatially unresolved. We also note that the scattered patterns surrounding the central source, including the circular patches in the northeast and southwest, are likely sidelobes in the SpM image, as they are consistent with the point-spread function pattern of the TM1 CLEAN image. For both the CLEAN and SpM images, the peak position measured by elliptical Gaussian fitting is $\rm \alpha(J2000) = 05^{h}55^{m}38^{s}.197$, $\rm \delta(J2000) = 2^{\circ}11'33''.59$. The fitting includes emission within the $\rm 10\sigma$ contour for the CLEAN image and within the $\rm \sim 10~\mu Jy~pixel^{-1}$ level for the SpM image. In this paper, we treat the continuum peak position as the central protostellar position, and all maps are centered on this position.

\begin{figure}
\plotone{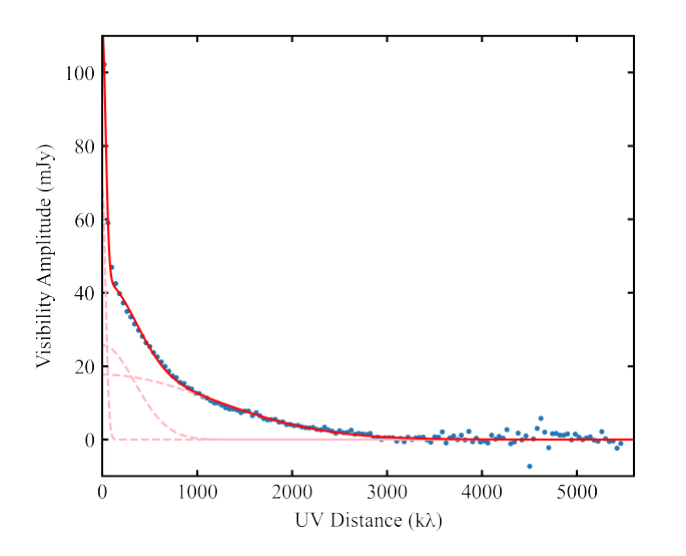}
\caption{Visibility amplitude vs. {\it uv} distance plot for the 1.3 mm continuum emission. The red solid curve denotes the sum of the three best-fit Gaussian components, while the light red dashed curves are the individual components. \label{fig:visibility}}
\end{figure}
Figure \ref{fig:visibility} shows the visibility plot derived by averaging all scans for each pair of antennas from the combined TM1+TM2 dataset. As shown by the red curve in Figure \ref{fig:visibility}, the visibility profile can be well-fitted by three Gaussian components centered at zero with size scales of $\rm 0''.066 \pm 0''.0021$ ($\rm 24 \pm 0.76\ au$), $\rm 0''.23 \pm 0''.010$ ($\rm 83 \pm 3.6 \ au$), and $\rm 2''.16 \pm 0''.073$ ($\rm 780 \pm 26 \ au$) in FWHM. The flux amplitudes given by the fitting are $\rm 17.7 \pm 0.93\ mJy$, $\rm 25.7 \pm 1.01\ mJy$, $\rm 68.6 \pm 1.80\ mJy$ for the smallest to largest components, respectively. The extremely compact continuum component suggests that a small circumstellar disk with a radius of $\rm \sim 12~au$, which is spatially unresolved in the image domain, has been formed in the early Class 0 stage. Meanwhile, the intermediate $\rm 0''.23$ ($\rm 83~au$) component could represent the inner part of the envelope where the material is being accreted into the disk. It is worth noting that there is no multiplicity at the current resolution. 

By assuming optical thinness and a { gas-to-dust} ratio of $100$, we can estimate the lower limit of the mass of each individual component with the following equation:
\begin{equation} \label{eqn:mass}
M_{\rm{comp}} = \frac{F_{\nu} { D}^{2}}{\kappa_{\nu} B_{\nu}(T_{\rm{dust}})}
\end{equation}
where $F_{\nu}$ is the total continuum flux density of the component derived from the visibility fitting, ${ D = \rm 360\ pc}$ is the distance to G204NE, $\kappa_{\nu}$ is the opacity coefficient, $B_{\nu}$ is the Planck function, and $T_{\rm dust}$ is the dust temperature. 
The {dust opacity per unit mass of gas and dust} was assumed to follow $\kappa_\nu \rm \sim 0.01\ {cm}^{2}\ g^{-1}$ at rest frequency $\nu \sim 233\rm \ GHz$ \citep{Ossenkopf1994}. 
{The $T_{\rm dust}$ was assumed to be $\rm 50~K$ for the most compact component and $\rm 25~K$ for both the intermediate and extended components, where the latter value corresponds approximately to the dust temperature derived from the SED fitting (Section \ref{subsec: SED}). We derived the three masses to be $\gtrsim 0.015$, $\gtrsim 0.049$, and $\gtrsim 0.13$ $M_{\odot}$ for the smallest to largest components, respectively.}

\subsection{Molecular Spatial Distribution} \label{subsec: molecular_lines}
\subsubsection{Dense Gas Tracers} 
\label{subsubsec:dense_gas}
\begin{figure*}[ht!]
\begin{center}
\includegraphics[width=0.95\textwidth]{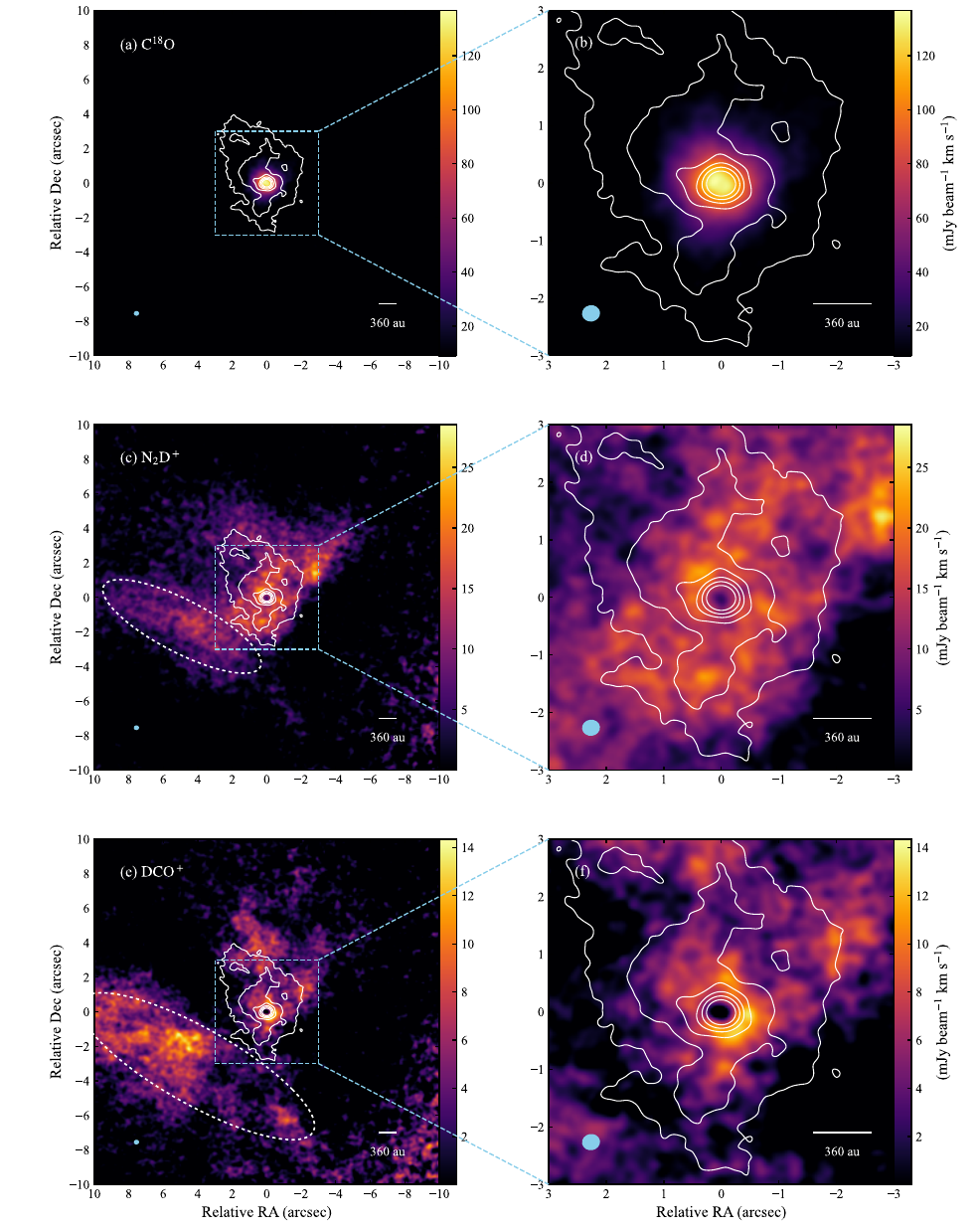}
\end{center}
\caption{Integrated intensity (moment 0) maps of $\rm C^{18}O$, $\rm N_2D^+$, and $\rm DCO^+$ as color and the TM1+TM2 continuum emission map as contour. In all panels, the contour levels are 5, 10, 20, 40, … $\times \sigma$, where $\sigma$ corresponds to $\rm 0.061~mJy~beam^{-1}$. The integrated observed velocity ranges are $v_{\rm obs}=$ $-0.3 - 3.1$, $0.8-2.3$, and $0.6-2.2$ $\rm km~s^{-1}$ for $\rm C^{18}O$, $\rm N_2D^+$, and $\rm DCO^+$, respectively. 
{The dashed ellipses in the panel (c) and (e) denote the spatially extended component.}
The bottom-left blue-filled ellipses in all panels show the {synthesized beams of the moment 0 maps}. \label{fig:moment0_dense}}
\end{figure*}
Figure \ref{fig:moment0_dense} shows the moment 0 maps of the $\rm C^{18}O$, $\rm N_2D^+$, and $\rm DCO^+$ lines as color and the TM1+TM2 continuum map as contour. The left panels are the wide field of views and the right panels are the close-up views of the central $\pm 3''$ region. Thanks to the higher-sensitivity TM2 dataset, the continuum emission can be seen elongated vertically at the $5\sigma$ contour level and shows an S-shaped pattern at the $10\sigma$ contour level; these features are not present in the TM1-only images (see Section \ref{subsec:continuum}). 
{The masses of the northern and southern arms surrounded by the 10 $\sigma$ contour are estimated to be 0.015 $M_{\odot}$ and 0.006 $M_{\odot}$, respectively, using Equation \ref{eqn:mass} and the same assumptions as the intermediate and extended components.}

The spatial distribution of the molecular lines demonstrates significant chemical stratification. Strong $\rm C^{18}O$ is detected toward the continuum peak, tracing the central compact component. The beam-deconvolved size of the C$^{18}$O emitting region derived from the elliptical Gaussian fitting is $1''.34 \times 1''.27$ ($\rm P.A. = -7.1^{\circ}$) in FWHM ($\rm \sim 470\ au$ in geometric mean). 
This suggests that $\rm C^{18}O$ is tracing the inner, warmer region of the envelope where $\rm CO$ is less depleted. Furthermore, the spatially compact $\rm C^{18}O$ indicates a small $\rm CO$ evaporation radius of $\sim$ 240 au around the embedded protostar.

In contrast, N$_2$D$^+$ and DCO$^+$ lines reveal the spatially extended distributions beyond the 5$\sigma$ contour of the continuum emission.
An additional, fainter component extending from the {south of the continuum source to the east} is seen in both N$_2$D$^+$ and DCO$^+$ {(dashed ellipses in Figure \ref{fig:moment0_dense}c and \ref{fig:moment0_dense}e, respectively)}.
This emission component likely traces the northern edge of the { curved emission feature linking two cores}  seen in the SCUBA-2 image.
Near the protostar, the spatial extent of $\rm DCO^+$ is in-between those of $\rm C^{18}O$ and $\rm N_2D^+$: $\rm N_2D^+$ is detected in most regions surrounding the protostar, while $\rm DCO^+$ mainly traces extensions to the north and to the northeast. 
Moreover, both $\rm N_{2}D^+$ and $\rm DCO^+$ exhibit a central hole and
show anti-correlation with $\rm C^{18}O$ and the continuum emission. This anti-correlation indicates that the $\rm N_2D^+$ and $\rm DCO^+$ lines are likely tracing the outer, colder regions where CO is frozen onto the dust and deuterium fractionation is enhanced. On the other hand, we detected no emission above 3$\sigma$ level in the $\rm DCN$ line. This also suggests the extended component surrounding the protostar is cold. Since the major formation path of DCN is from CH$_2$D$^+$ in the temperature range of 30--70 K (\citealp{Millar1989, Turner2001}), DCN is not formed efficiently in the extended envelope where the temperature is below the CO submlimation temperature.

\subsubsection{Outflow and Jet Tracers} \label{subsubsec:outflow_tracers}
\begin{figure*}[ht!]
\begin{center}
\includegraphics[width=\textwidth]{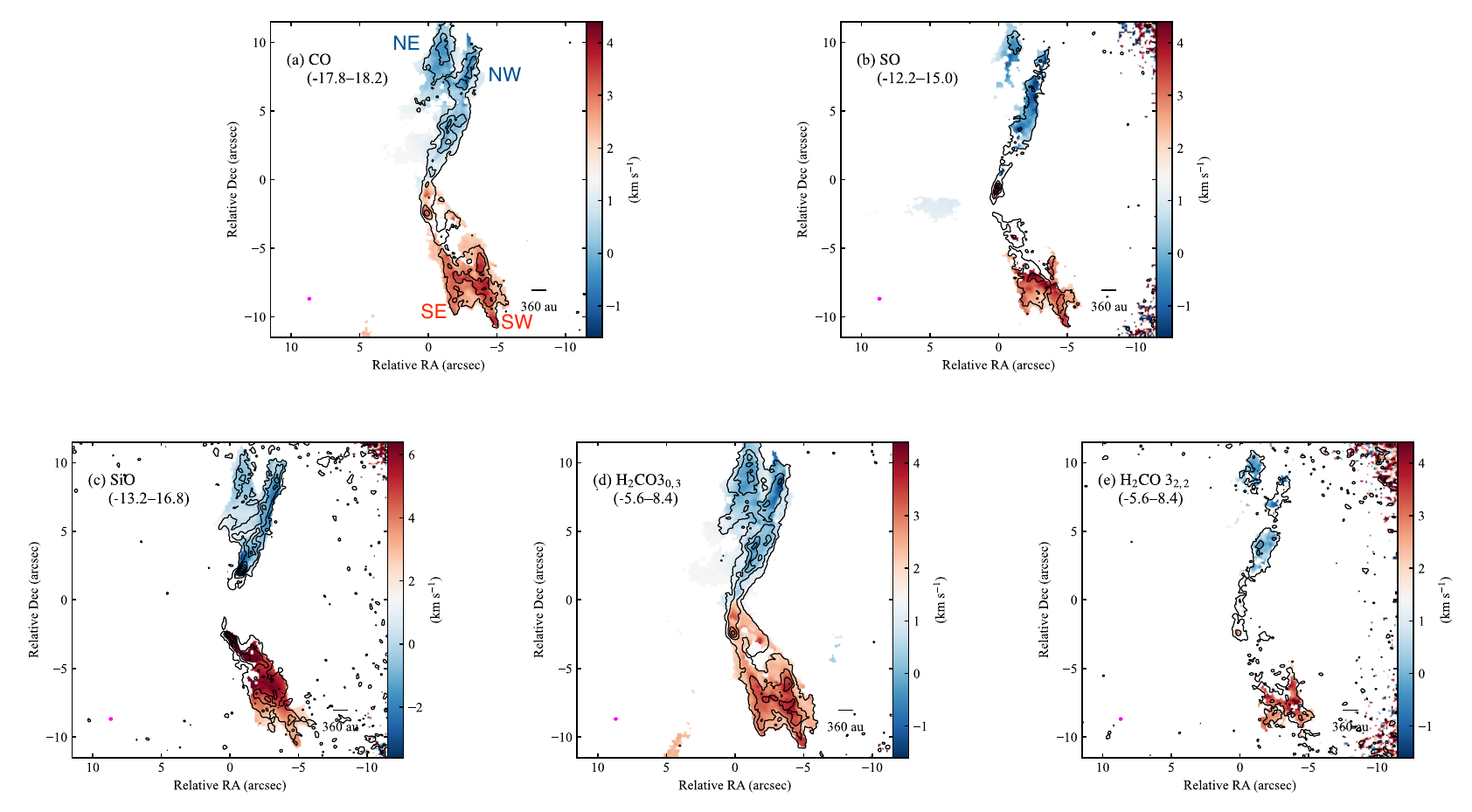}
\end{center}
\caption{Intensity-weighted average velocity (moment 1) maps of the outflow tracers as color maps, overlaid with the integrated intensity (moment 0) maps drawn in contours. The contour levels are $3$, $6$, $9$, ... $\times \sigma$, where $\sigma$ corresponds to $4.0$, $2.7$, $2.1$, $2.1$, and $2.0$ $\rm mJy~beam^{-1}~km~s^{-1}$ for (a)-(e), respectively. Pixels with intensities below $5\sigma$ are excluded in the moment 1 maps. The velocity offset ($v_{\rm obs} - v_{\rm sys}$) ranges of the maps in units of $\rm km~s^{-1}$ are as indicated in parentheses. Magenta-filled ellipse in the bottom left corner of each panel shows the  synthesized beam.}
\label{fig:outflow_mom0}
\end{figure*}

Figure \ref{fig:outflow_mom0} shows the moment 0 (contour) and moment 1 { (color)} maps of the $\rm CO$, $\rm SiO$, $\rm SO$, and two transitions of $\rm H_2CO$ (hereafter referred to as $\rm H_2CO\ 3_{03}$ and $\rm H_2CO\ 3_{22}$). 
The integrated velocity range is $\pm$20 km s$^{-1}$ from the cloud systemic velocity, $v_{\rm{sys}}=$1.4 km s$^{-1}$ (see Section \ref{subsec: kinematics}). Highly-collimated outflow along the north-south direction is observed in all these molecular lines. As seen in the moment 1 maps, the northern lobe is blueshifted and the southern lobe is redshifted.
The position angle of the blueshifted lobe is $\sim -15^{\circ}$, while that of the redshifted lobe is $\sim 200^{\circ}$. As shown in the CO channel maps (Figures \ref{fig:CO_channel_map} and \ref{fig:CO_channel_map_lowvel}), the eastern wall of the northern lobe having a position angle of $\sim -5^{\circ}$ (yellow dashed line in Figures \ref{fig:CO_channel_map} and \ref{fig:CO_channel_map_lowvel}) appears only in the low-velocity channels of $-3 < v_{\rm obs} - v_{\rm sys} < -1 $ km s$^{-1}$. 
In the velocity range of $-6 < v_{\rm obs} - v_{\rm sys} < -3$ km s$^{-1}$, emission comes from the western wall with a position angle of $\sim -22^{\circ}$ (blue dashed line in Figures \ref{fig:CO_channel_map} and \ref{fig:CO_channel_map_lowvel}). 
Such a velocity asymmetry is also seen in the southern lobe; the western wall that is seen in the velocity channels of $v_{\rm obs} - v_{\rm sys} < 4$ km s$^{-1}$ disappears in the higher velocity channels.
The velocity asymmetry in the northern lobe is commonly observed in all molecular lines, while that in the southern lobe is less significant in the H$_2$CO lines. 
The farthest tips of the blueshifted and redshifted lobes exhibit a pair of bow-like patterns labeled NE and NW in the northern tip and SE and SW in the southern tip.
{The bow-like patterns, which are most prominently observed in the CO and $\rm H_2CO~3_{0,3}$ (Figures \ref{fig:outflow_mom0}(a) and (d)), are likely attributable to interactions with the surrounding gas.} 
Both northern and southern lobes are spatially extended up to $\gtrsim 11.5''$ ($\rm 4140~au$).

The CO channel map in the low velocity range (Figure \ref{fig:CO_channel_map_lowvel}) shows that the CO emission extends to the NE in the {redshifted} velocity range.
{ Since the} low-velocity redshifted component extends along the axis of the southern lobe,
this {NE} component could be another outflow lobe.
This component is observed only in CO.
The observed low radial velocity is attributed either to the inclination angle being nearly parallel to the plane of the sky or to the intrinsically low velocity.
The latter case of intrinsically low velocity outflow is expected if this component is driven when the central source was in the first hydrostatic core stage.
However, the nature of this low-velocity redshifted component is not well understood because the blueshifted counterpart of this component is unclear.
The CO emission in the { low-velocity blueshifted component ({ Figure \ref{fig:CO_channel_map_lowvel})} extends towards the southeast; however, the spatial distribution of the low-velocity blueshifted component is amorphous and does not point to the central source.
Figure \ref{fig:CO_LVC} { compares the spatial distributions of the continuum emission and the low-velocity outflow components.
It} illustrates that the S-shaped feature seen in the continuum emission outlines the boundaries of the outflow lobes: the northern arm extends between the base of the NE component and the blueshifted lobe, and the southern arm aligns with the eastern boundary of the redshifted lobe.

\begin{figure*}[ht!] 
\epsscale{1.0}
\plotone{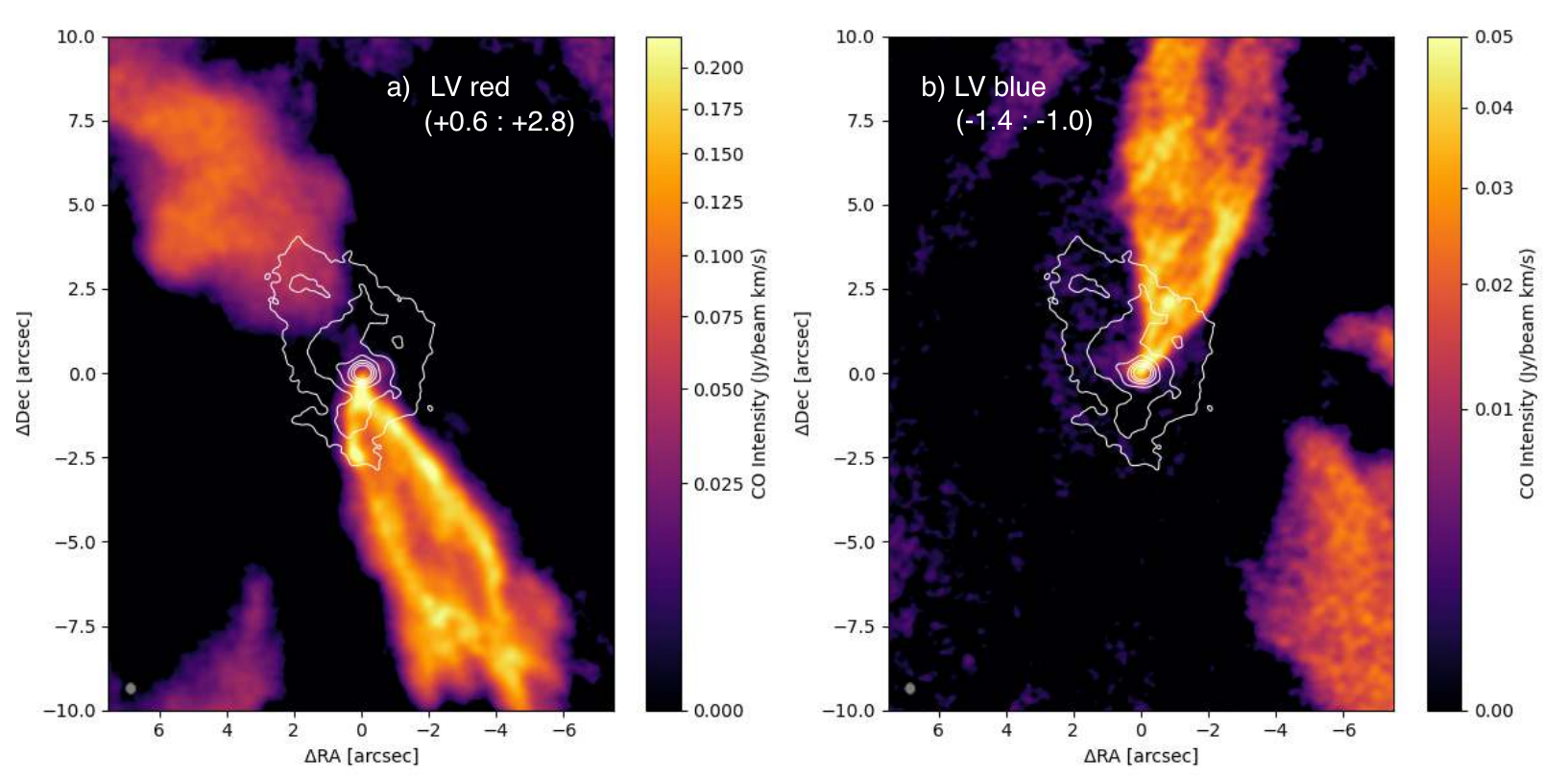}
\caption{{Low-velocity CO (a) redshifted and (b) blueshifted components (color) overlaid with the TM1+TM2 continuum emission map as contour. 
Integrated velocity offset ranges are (a) from 0.6 to 2.8~km~s$^{-1}$ and (b) from $-$1.4 to $-$1.0 km s$^{-1}$.
Contour levels are same as those of Fig. \ref{fig:moment0_dense}.}
{ Grey ellipse denotes at the bottom-left the synthesized beam of the CO image.}}
\label{fig:CO_LVC}
\end{figure*}

\begin{figure*}[ht!] 
\centering
\includegraphics[width=\textwidth]{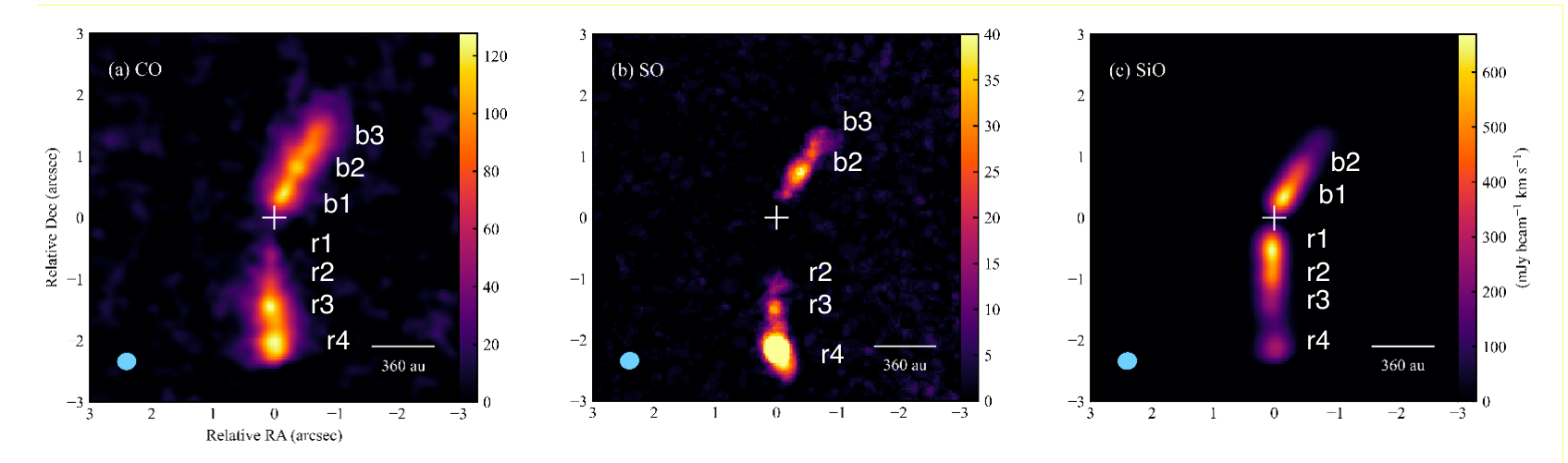}
\caption{Moment 0 maps of (a) $\rm CO$, (b) $\rm SO$, and (c) $\rm SiO$ integrated with the velocity offset ranges of $v_{\rm obs} - v_{\rm sys}$ are from $-38.4$ to $-21.4$ km s$^{-1}$ and from $22.0$ to $36.6$ km s$^{-1}$. The bottom-left blue-filled ellipse denotes the synthesized beam.}
\label{fig:jet_mom0}
\end{figure*}

The CO, SiO, and SO lines exhibit higher velocity emission at $|v_{\rm{obs}}-v_{\rm{sys}}| \gtrsim 18\rm ~km~s^{-1}$.
As shown in Figures \ref{fig:jet_mom0}(a)-(c), this extremely high-velocity component is spatially compact and highly collimated.
It is likely that this extremely high-velocity emission comes from the jet.
The blueshifted jet in the north and redshifted jet in the south have position angles of $\sim -30^{\circ}$ and $\sim 90^{\circ}$, respectively. The jets are extended up to $\sim 2''$ ($\rm 720~au$).
Several knot-like structures {with a separation of $\sim$0\farcs5} are observed along the jet axes.
The SiO emission is brightest in the innermost knot pair labeled b1 and r1, and becomes fainter in the outer knots.
In contrast to SiO, the SO emission tends to be bright in the outer knots, especially at the r4 knot, and barely seen in the innermost knots.
All of the knots are seen in the CO, although the innermost r1 knot is fainter than its counterpart b1 knot.

\begin{figure}[ht!] 
\epsscale{1.2}
\plotone{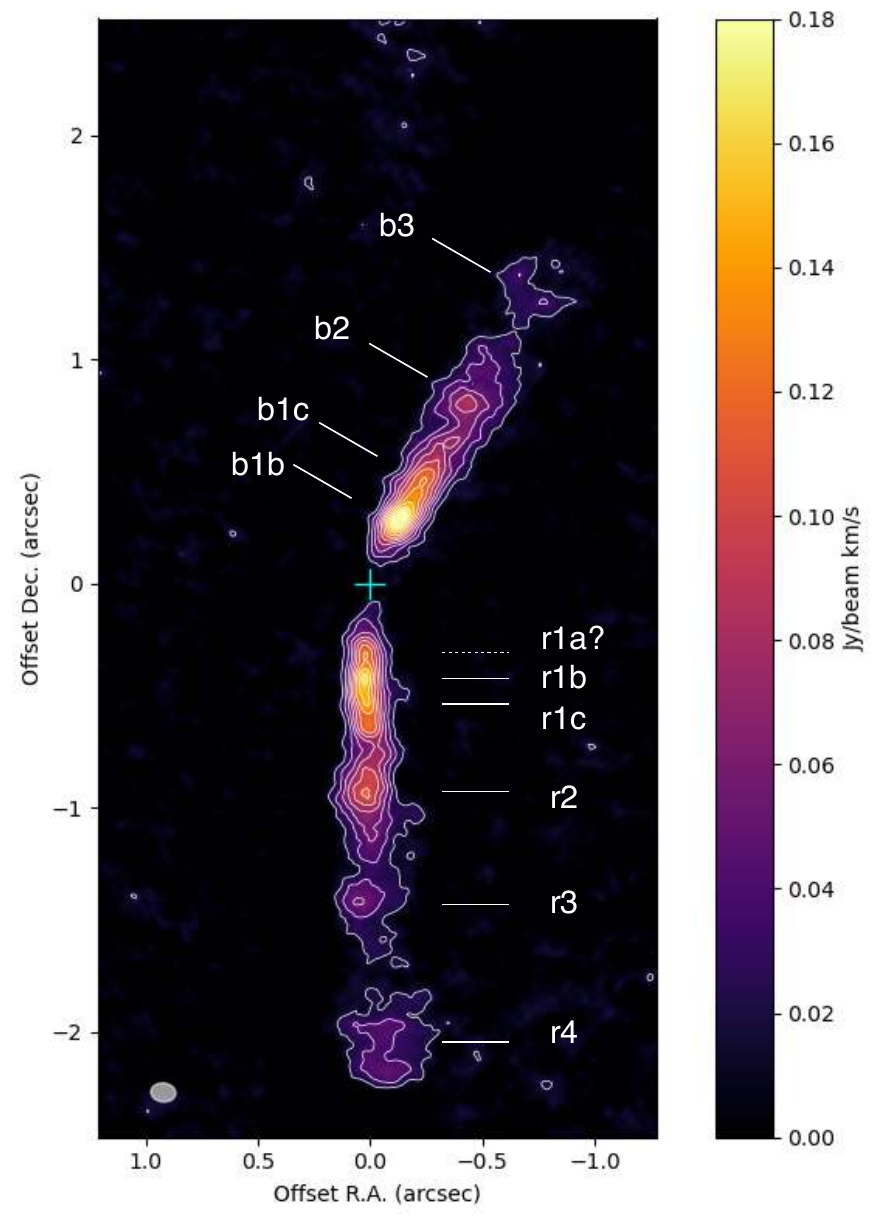}
\caption{{High-velocity $\rm SiO$ emission observed with TM1. 
The integrated velocity ranges are same as those of Figure \ref{fig:jet_mom0}.
The bottom-left blue-filled ellipse in each panel denotes the synthesized beam.}}
\label{fig:SiO_TM1}
\end{figure}

{The 0\farcs1 resolution image of the SiO jet observed with TM1 (Figure \ref{fig:SiO_TM1}) reveals its detailed structure.
At high resolution, the innermost pair of knots, r1 and b1, are elongated along the jet axes.
The transverse widths of these knots after deconvolved with the beam are $\sim$21 au in r1 and $\sim$27 au in b1.
In addition to the brightest peaks labeled r1b and b1b, there are additional emission peaks r1a, r1c, and b1c.
The elongated structure with sub-knots observed in r1 and b1 is quite similar to that of HH211 \citep{Lee2009,Jhan2021}.}

\section{Analysis} \label{sec:analysis}
\subsection{Molecular Abundances}
\label{subsec:molecular_abundances}

\begin{figure*}[ht!] 
\includegraphics[width=\textwidth]{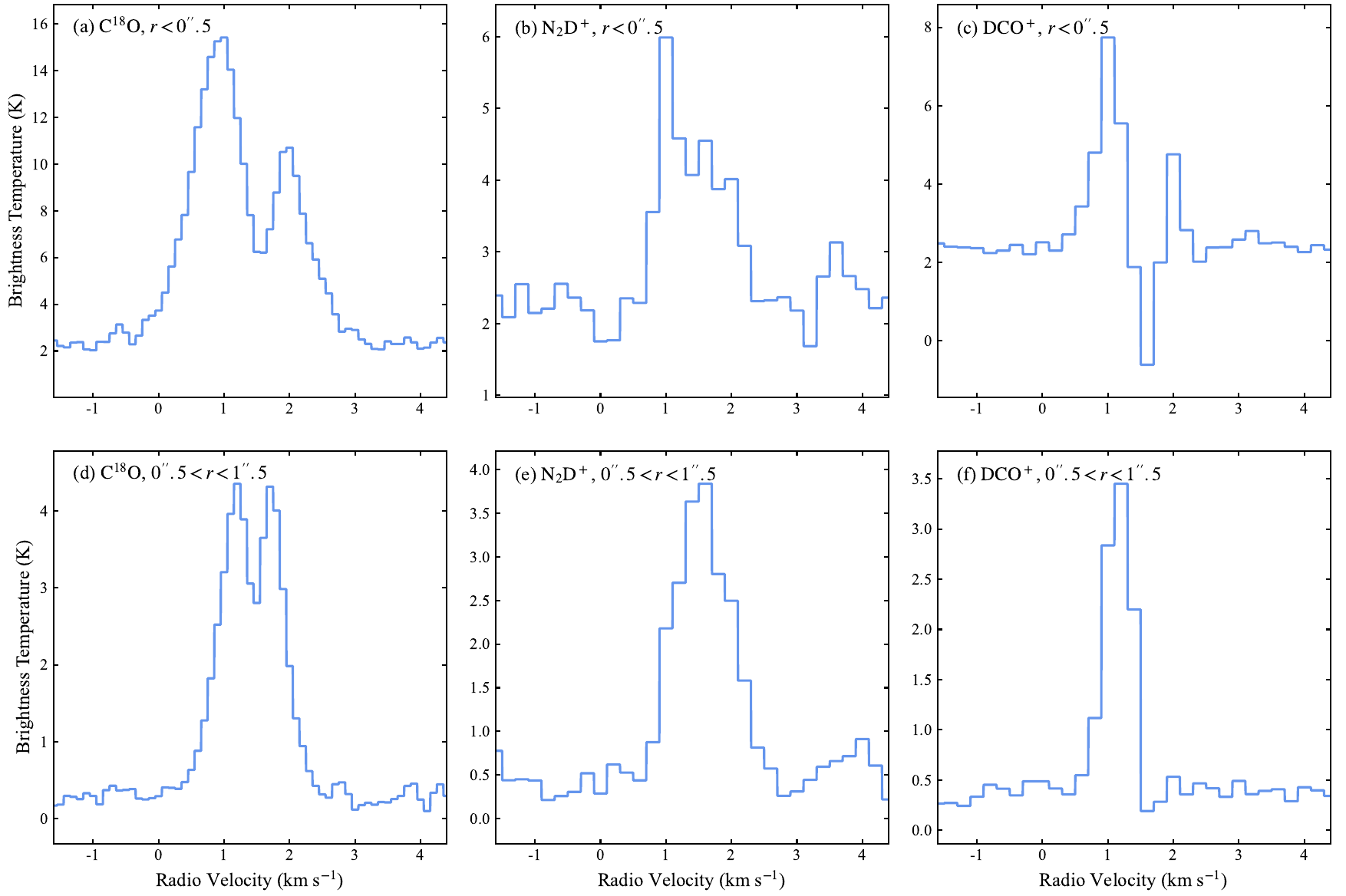}
\caption{Spectral profiles of $\rm C^{18}O$, $\rm N_2D^+$, and $\rm DCO^+$ before continuum subtraction. 
{The top panels (a)-(c) show the profiles obtained from a circular aperture with a radius of $r < 0''.5$ centered at the continuum peak position, while the bottom panels (d)-(f) show those obtained from the annular region with $0''.5 < r < 1''.5$.}}
\label{fig:spectra}
\end{figure*}

To estimate the fractional abundances ($N({\rm{X}})/N(\rm{H_2})$) of $\rm C^{18}O$, $\rm N_2D^+$, and $\rm DCO^+$, we first individually derived the molecular column densities using different assumptions. Then, we divided the molecular column densities by the $\rm H_2$ column density derived from the continuum emission. 
If we assume that the continuum emission is optically thin, the $\rm H_2$ column density can be related to the continuum flux density as in the following equation: 
\begin{equation}
N(\rm{H_2}) = \frac{S_\nu}{\kappa_\nu \ B_\nu(T_{\rm{dust}})\ m_p\ \mu_{\rm H_2}\ \Omega_{\rm{beam}}} 
\end{equation}
where $S_\nu$ is the continuum flux density, $\kappa_\nu$ $\sim 0.01\rm \ cm^2\ g^{-1}$ is the opacity coefficient at rest frequency $\nu \sim \rm 233\ GHz$ \citep{Ossenkopf1994}, $B_\nu(T_{\rm dust})$ is the Planck function at dust temperature $T_{\rm dust}$, $\mu_{\rm H_2} \rm =2.8$ is the mean molecular weight, and $\Omega_{\rm beam}$ is the beam solid angle. By assuming $T_{\rm dust} = 50~ \rm K$ in the $r < 0''.5$ inner region {and $T_{\rm dust} = 25~ \rm K$ in the outer $0''.5 < r < 1''.5$ region} 
, we derived the average $\rm H_2$ column density {to be $\sim 8.0 \times 10^{23} \rm~cm^{-2}$ for the inner region and $\sim 1.8 \times 10^{23}~\rm cm^{-2}$ for the outer region}.
{Since the optically thin assumption is not valid for the innermost region, the H$_2$ column density at $r < 0''.5$ derived here is the lower limit.}

\subsubsection{C$^{18}$O Abundance} 
\label{subsec: C18O_abundance}

We derived the C$^{18}$O column density by assuming the local thermodynamic equilibrium { (LTE)} condition.
Since the spatial distribution of the C$^{18}$O emission correlates well with that of the continuum emission, we assumed that the line emission comes from the same region as the continuum, and that the line excitation temperature is same as the dust temperature. 
The continuum intensity $I_{\rm cont}$ can be related to the dust optical depth $\tau_{d}$ using the radiative transfer equation as 
\begin{equation}
I_{\rm{cont}} = (B_{\nu}(T) - B_{\nu}(T_{\rm{bg}}))(1 - e^{-\tau_{d}})
\end{equation}
where $B_{v}$, $T$, $T_{bg} = { 2.7}~\rm K$ are the Planck function, the excitation/dust temperature, and the cosmic microwave background temperature, respectively. The line intensity $I_{\rm{line}}$ before continuum subtraction is given by
\begin{equation}
I_{\rm{line}} = (B_{\nu}(T)-B_{\nu}(T_{\rm{bg}}))(1-e^{-(\tau_{g}+\tau_{d})})
\end{equation}
where $\rm \tau_{g}$ is the gas optical depth. 

{Assuming $T=\rm 50~K$ for the inner $r<0''.5$ region and $T=\rm 25~K$ for the outer $0''.5<r<1''.5$ region, we calculated the gas optical depth at each channel from observed intensities $>5\sigma$ and estimated the integrated gas optical depth, $\int \tau_gdv$, by summing over the values at each channel. Then, we calculated the $\rm C^{18}O$ column densities from the integrated gas optical depths} using the following equation from \cite{Mangum2017}:
\begin{equation} \label{eqn:coldens}
N_{\rm tot} = 
\frac{3h}{8\pi^{3}|\mu_{lu}|^2}
\frac{Q_{\rm rot}}{g_{u}}
\exp(\frac{E_{u}}{kT_{\rm ex}}) \times
[\exp(\frac{h\nu}{kT_{\rm ex}})]^{-1}
\int \tau_{g} dv
\end{equation}
where $h$ is the Planck constant, $k$ is the Boltzmann constant, $|\mu_{lu}|$ is the dipole matrix moment, $Q_{\rm rot}$ is the rotational partition function, $g_{u}$ is the upper energy level's degeneracy, $E_u$ is the energy of the upper transition level, $\nu$ is the $\rm C^{18}O$ rest frequency, and $T_{\rm{ex}}$ is the excitation temperature, {which we assumed to be the same as the dust temperature}. The rotational partition function for linear molecules was assumed to follow $Q_{\rm{rot}} \sim \frac{kT_{\rm{ex}}}{hB_0} + \frac{1}{3}$ using the series approximation method \citep{Mangum2017}, where $B_0$ is the rotational constant. The values of $|\mu_{lu}|$ and $B_0$ were obtained from the { Jet Propulsion Laboratory (JPL) Molecular Spectroscopy database \citep{Pic98}} through the Splatalogue Database for Astronomical Spectroscopy (url: \url{splatalogue.online}). 

{We derived the $\rm C^{18}O$ column density to be $\sim 1.6 \times 10^{16}$ cm$^{-2}$ in the inner $r<0''.5$ region and $\sim 1.0 \times 10^{13}$ cm$^{-2}$ in the outer $0''.5<r<1''.5$ region. The $\rm C^{18}O$ abundance is thus $\sim 2.0 \times 10^{-8}$ and $\sim 5.3 \times 10^{-11}$ in the inner and outer regions, respectively. 
Despite the higher abundance of C$^{18}$O in the inner region, its absolute value is approximately an order of magnitude lower than the commonly accepted C$^{18}$O abundance for the interstellar medium, which is (1.7--2)$\times$10$^{-7}$ (\citealp{Wannier1980, Frerking1982})}. 
It should be noted that the C$^{18}$O line profiles in the inner and outer regions have two peaks with a dip in $V_{\rm LSR}{\sim}$~1.6 km s$^{-1}$.
The observed dip is likely attributable to absorption from the spatially extended foreground component. Consequently, the C$^{18}$O column densities previously derived may be underestimated. Nevertheless, it is improbable that this absorption feature reduces the column densities by an entire order of magnitude.

\subsubsection{$N_2D^+$ Abundance} \label{subsec:N2Dp_abundance}
By assuming {LTE,} optical thinness, {and} $T_{\rm{ex}} = 15\ \rm{K}$, we derived the lower limits of the $\rm N_2D^+$ column density using the following equation for linear molecules from \cite{Mangum2017}: 
\begin{equation}
\label{eqn:optically_thin}
    N^{\rm thin}_{\rm tot} = 
    \frac{3h}{8\pi^3 |\mu_{lu}|^2}
    \frac{Q_{\rm rot}}{g_u}
    \frac{\exp(\frac{E_u}{kT_{\rm ex}})}{\exp(\frac{E_u}{kT_{\rm ex}})-1}
    \frac{\int T_B dv}{J_\nu(T_{\rm ex})-J_\nu(T_{\rm bg})}
\end{equation}
where the symbols are the same as defined in Equation \ref{eqn:coldens}, and $J_\nu$ and $\int T_{B} dv$ are the Rayleigh-Jeans Equivalent Temperature and the integrated brightness temperatures derived from the moment 0 map before continuum subtraction, respectively.
We assume $T_{\rm bg}$ to be 2.7 K for the outer region at $0''.5 < r < 1''.5$ and $J_{\nu}(T_{bg})$ to be $2.2\rm ~K$ for the inner $r < 0''.5$ region, {the latter of} which is the continuum level of the spectrum  (Figure \ref{fig:spectra}(b)).

The derived N$_2$D$^+$ column densities are ${\sim}~ 6.4 \times 10^{11}$ cm$^{-2}$ in the inner $r<0''.5$ region and ${\sim}~2.1 \times 10^{11}$ cm$^{-2}$ in the outer $0''.5<r<1''.5$ region.
The N$_2$D$^+$ abundance is thus ${\sim}~8.0 \times 10^{-13}$ in the inner region and ${\sim} 1.1 \times 10^{-12}$ in the outer region.
{ The uncertainties in the abundance values, which stem from the uncertainties linked to the $\rm H_2$ column densities and the rms noise level of the spectra, are estimated to be ${\sim} (3-4) \times 10^{-14}$.
Therefore, the observed reduction in the N$_2$D$^+$ abundance within the inner region is noteworthy.}
The drop in $\rm N_2D^+$ abundance near the protostellar position is consistent with the explanation that the early-phase protostar has started to heat the gas and evaporate $\rm CO$ from the dust, which then reacts with $\rm N_2D^+$ to form $\rm DCO^+$.

\subsubsection{$DCO^{+}$ Abundance} 
We used the same assumptions of {LTE}, optical thinness, and $T_{\rm{ex}} = 15\ \rm{K}$ to derive the $\rm DCO^+$ column densities. 
{ We also used the same assumption for the background temperature, i.e. $T_{\rm bg} =$ 2.7 K for the outer region and $J_{\nu} (T_{\rm bg}) = $ 2.2 K for the inner region.}
In certain channels, the $\rm DCO^{+}$ line shows a dip below the continuum level.
Since the brightness temperature at the bottom of the dip is close to zero, the dip can likely be attributed to the foreground low-excitation component.
(Figure \ref{fig:spectra}(c)). 
Therefore, we excluded channels where the absorption is shown  (i.e., $v_{\rm{obs}} = 1.4-1.8\rm ~km~s^{-1}$) in our calculations. 

The column densities of DCO$^+$ in the inner $r < 0''.5$ region and the outer $0''.5 < r < 1''.5$ region are ${\sim}~3.7 \times 10^{15}~\rm cm^{-2}$ and ${\sim}~8.8 \times 10^{14}\rm ~cm^{-2}$, respectively. 
The derived $\rm DCO^{+}$ abundances in the inner and outer regions are not significantly different, ${\sim}~4.6  \times 10^{-9}$ and ${\sim}~4.8 \times 10^{-9}$, respectively.
The derived column densities and abundance values of DCO$^+$ presented in this study represent the lower limits for both the inner and outer regions, as the velocity channels exhibiting absorption features were omitted from the column density calculations of DCO$^+$.

\subsection{Envelope Kinematics}
\label{subsec: kinematics}

\begin{figure*}[ht!]
\includegraphics[width=\textwidth]{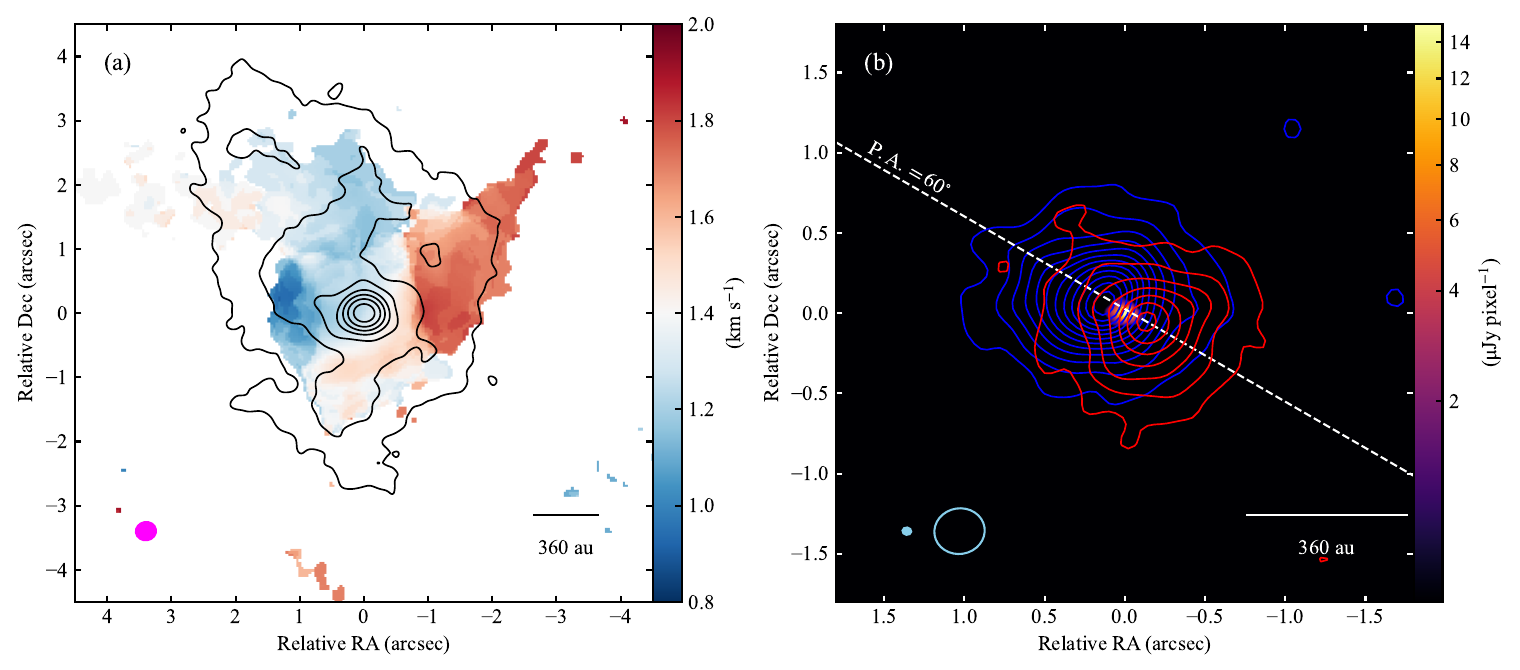}
\caption{(a) Moment 1 map of C$^{18}$O (color) overlaid with the TM1+TM2 continuum map (contour). 
Contour levels are 5, 10, 20, 40, … $\times \sigma$ ($\sigma = \rm 0.061~mJy~beam^{-1}$).
Moment 1 map was made with $\rm C^{18}O$ emissions below $5\sigma$ ($\sigma=2.5~\rm mJy~beam^{-1}$) excluded.
Magenta-filled ellipse denotes the synthesized beam.
{(b) Moment 0 maps of the high-velocity C$^{18}$O emission components overlaid on the SpM continuum image.
The integrated velocity ranges of the blueshifted and redshifted components are from $-$0.3 to $+$0.6 km s$^{-1}$ and from $+$2.2 to $+$3.0 km s$^{-1}$, respectively.
White dashed line denotes the direction of the velocity gradient, P.A. = 60$^{\circ}$.
Open and filled ellipses are the beam of the C$^{18}$O image and the effective resolution of the SpM continuum image, respectively.}
\label{fig:kinematics}}
\end{figure*}

Figure \ref{fig:kinematics}(a) shows the $\rm C^{18}O$ intensity-weighted average velocity (moment 1) map as the color map and the continuum as the contour map. The complete channel maps are also shown in Figure \ref{fig:C18O_channel_map} in Appendix \ref{sec:channel_maps}. The $\rm C^{18}O$ velocity distribution exhibits a blueshifted component to the east and a redshifted component to the west.
Since this velocity gradient is approximately perpendicular to the outflow axes, it is interpreted as a rotational motion of the envelope.
In the channel map (Figure \ref{fig:C18O_channel_map}), the emission with large velocity offsets  appears in the vicinity of the central source, suggesting that the inner region is rotating faster.
The spatial distributions of the high velocity components presented in Figure \ref{fig:kinematics}(b) reveals that the blueshifted component is located in the NE and the redshifted component is in the SW of the central source. 
The orientation of the velocity gradient has been determined to be P.A. $\sim$ 60$^{\circ}$ from the positions of the emission centroids at channels with $|v_{\rm obs}-v_{\rm sys}|~>$ 0.8 km s$^{-1}$.
Therefore, we assume that the envelope is rotating around the axis with a position angle of ${\sim}-$30$^{\circ}$, which is parallel to the axis of the blueshifted jet.

\begin{figure*}[ht!]
\includegraphics[width=\textwidth]{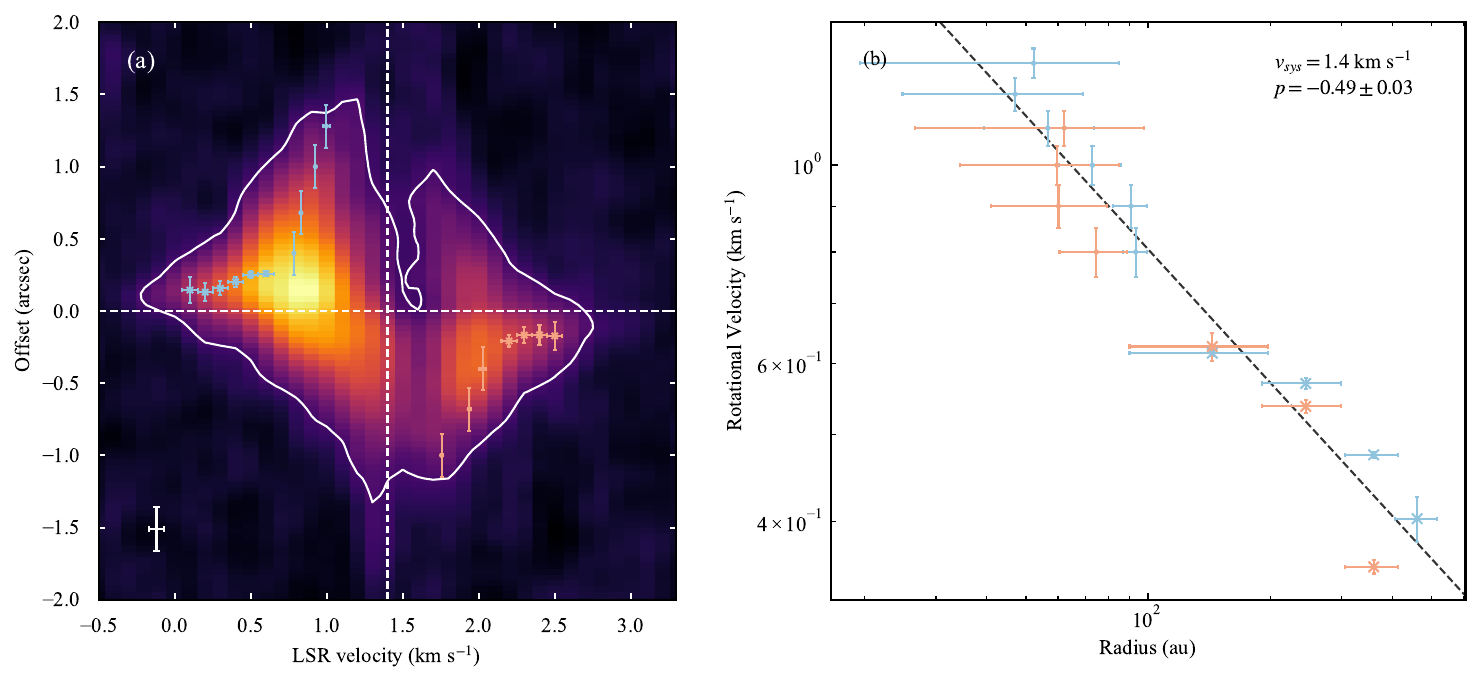}
\caption{{(a) Position-velocity (PV) diagram of C$^{18}$O along the P.A. = 60$^{\circ}$ line passing through the position of the central source.
White contour denotes the 5$\sigma$ level.
Blue and red points with error bars are the measured rotational velocities of the NE side (blueshifted) and the SW side (redshifted), respectively.
Horizontal and vertical dashed lines are the position of the central source and the systemic velocity, respectively.
The horizontal and vertical error bars in the bottom left corner denote the velocity resolution of the data cube and the beam dimension along the position angle of the PV slice.
(b) Log-log plot of the measured rotational velocity as a function of the projected radius.
Blue and orange dots with error bars are the intensity weighted positions at given velocities.
Blue and orange crosses with error bars are the centroid velocities of the blue and red components, respectively, derived from the two-component Gaussian fitting.
}}
\label{fig:C18OPV}
\end{figure*}

Figure \ref{fig:C18OPV}(a) shows the position-velocity (PV) diagram along the velocity gradient, i.e. P.A.~=~60$^{\circ}$.
As expected from the line profile (Figure \ref{fig:spectra}(a)) the blueshifted component in the second quadrant is brighter than the redshifted component in the fourth quadrant.
The P-V diagram exhibits the spin-up feature in which the rotational velocity increases toward the central source. 
The low-velocity emission extends on either side of the central source, and exhibits two triangular structures with their tips point to the highest blue- and redshifted velocities.
Such a two-triangular { feature} has often been observed in protostellar envelopes, and has been interpreted as an infalling motion \citep[e.g.][]{Ohashi1997}.

The spin-up feature of the envelope has been examined using the rotational velocities as a function of radius.
Following \citet{Yen2013}, we adopted different methods to the inner and outer part; the intensity profile at each velocity channel for the inner high-velocity part with $|v_{\rm obs}-v_{\rm sys}|~>$ 0.8 km s$^{-1}$, and the spectral profile at a given position for the outer low-velocity part.
We derived an intensity-weighted position for the inner part.
For the outer part, each spectral profile was fitted with double-Gaussian, and the velocity of the bluer component is adopted in the NE side, and that of the redder component is adopted in the SW side.
The rotational velocities derived in these methods are plotted on the P-V diagram (Figure \ref{fig:C18OPV}(a)).

Figure \ref{fig:C18OPV}(b) presents the power-law dependence of the rotational profile.
A least-square fitting, using a power law ($\propto~r^{\alpha}$), to all the data points provides an index $\alpha$ of $-$0.49$\pm$0.03, which is expected for the case of Keplerian rotation, $-$0.5, and different from $-$1 for the infall with angular momentum conservation.
In the case of Keplerian rotation, the central mass { enclosed within a radius of $\sim$ 45 au }is estimated to be (0.071~$\pm$~0.001)/${\rm cos}^2 ~i$, where $i$ the inclination angle of the rotational axis from the plane of the sky.
Assuming that the rotational axis is parallel to the axis of the outflow ($i~{\sim}~40^{\circ}$, see Section \ref{subsubsec:outflow_modeling}) or jet ($i~{\sim} ~30^{\circ}$, see Section \ref{subsubsec:jet_modeling}), the { enclosed} mass is derived to be 0.121$\pm$0.002 $M_{\odot}$ and 0.095$\pm$0.001 $M_{\odot}$, respectively.

The inner and outer parts exhibit similar slope;
the slope does not steepen in the outer region regardless that the double-triangle feature in the P-V diagram implies the presence of infall.
\citet{Yen2013} investigated the rotational profiles of the model images of the infalling and rotating envelopes, and compared the genuine rotational profiles and the ones derived from the P-V diagrams of the simulated observations.
They found that the rotational profiles of the simulated observations tend to have shallower slopes than those of the input values.
The rotational profile of $\propto r^{-0.5}$ is also derived from the three-dimensional infalling and rotating envelope models of \citet{Mori2024}.
As discussed in \citet{Yen2013} and \citet{Mori2024}, the observed shallower rotational profiles { can be attributed to} the contribution of the radially infalling gas to the line-of-sight velocities.
{ Thus}, the rotational profile of $\propto r^{-0.5}$ observed in the envelope of G204NE is not { necessarily a result of} Keplerian rotation.

Therefore, we analyzed the observed P-V diagram using a simple toy model of a rotating and infalling envelope. For simplicity, we assumed a flattened envelope structure { with a radius of $R_{\rm env} = \rm 540~au$, which corresponds approximately to the largest positional offset traced by the $5\sigma$ contour in the P-V diagram (Figure \ref{fig:C18OPV}a)}.
Following \cite{Sakai2014}, we { assumed that the} flared envelope has { bipolar conical} cavities of full opening angles $\theta$.
{ Here,} $\theta = 180^{\circ}$ represents the case of a spatially thin disk, while $\theta = 0^{\circ}$ represents that of a sphere with no cavity. For this model, we assume that both the total energy and the angular momentum are conserved in the infalling material. 
We { also} assumed a large opening angle of $\theta \rm \geq 130^{\circ}$ to allow for a small angle approximation error of $\rm \lesssim 10\%$. Thus, the magnitudes of the infall velocity $v_{\rm{inf}}$ and the rotational velocity $v_{\rm{rot}}$ can be related to the specific angular momentum $l$, the radius from central protostar $r_{\rm{cen}}$, the shortest distance to the axis of rotation $r_{\rm{ax}}$, and the stellar mass $M_{\star}$ as:
\begin{equation}
v_{\rm{inf}} = \sqrt{\frac{2GM_{\star}}{r_{\rm{cen}}} - (\frac{l}{r_{\rm{ax}}})^2}
\end{equation}
and
\begin{equation}
v_{\rm{rot}} = \frac{l}{r_{\rm{ax}}}
\end{equation}
where $G$ is the universal gravitational constant. The radius of centrifugal barrier $r_{\rm CB}$ (where all kinetic energy is converted into rotational motion) can hence be given by:
\begin{equation}
\label{eqn:CB}
r_{\rm CB}= \frac{l^2}{2GM_{\star}}
\end{equation}

\begin{figure*}[ht!]
\includegraphics[width=\textwidth]{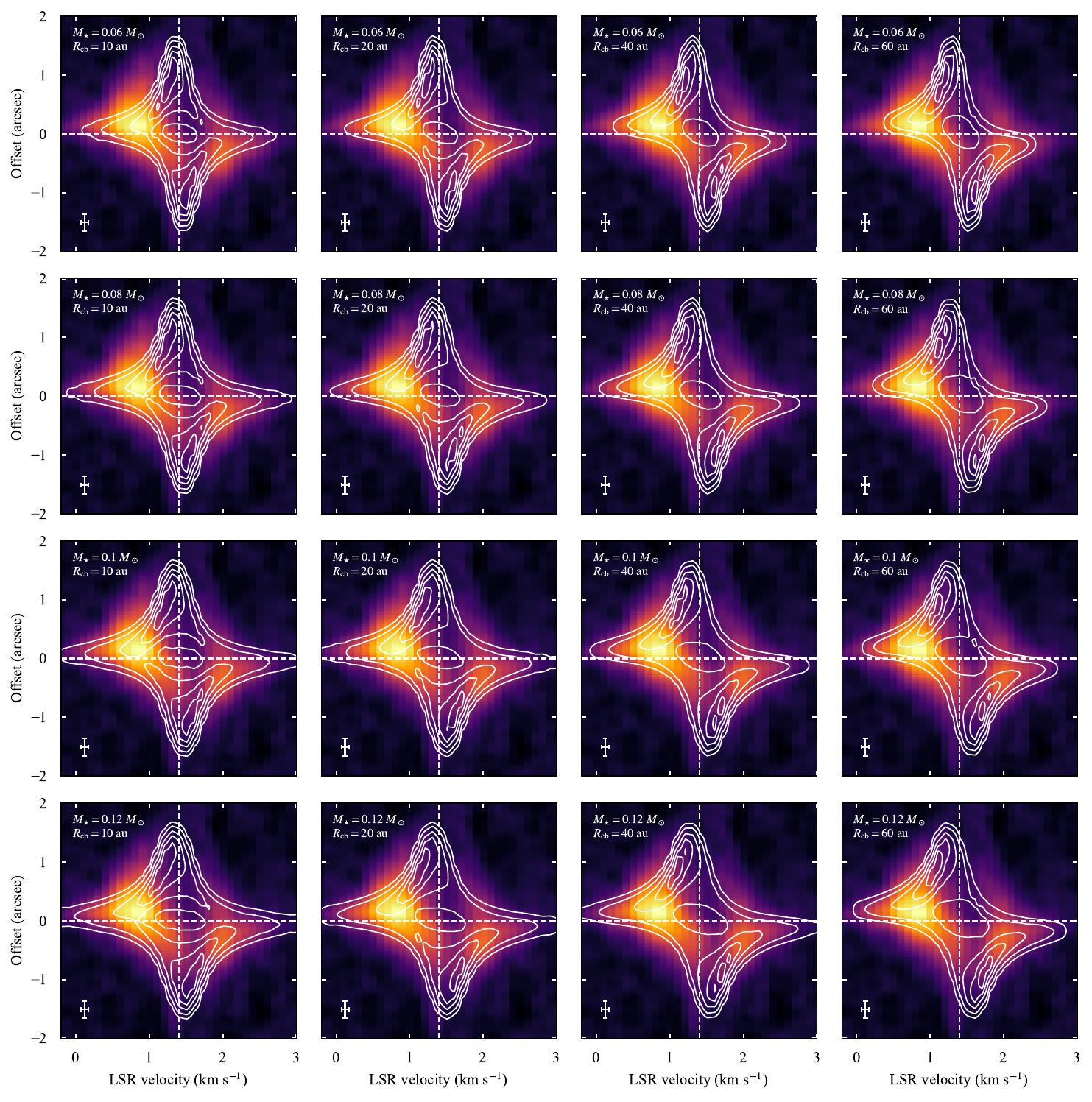}
\caption{{Position–velocity (PV) diagram of $\rm C^{18}O$ emission along $\rm P.A = 60^{\circ}$, overlaid with contours of the toy model for the specified parameters. Each row corresponds to a different stellar mass (increasing from top to bottom), and each column corresponds to a different centrifugal { barrier} (increasing from left to right). The contour levels are $0.09$, $0.2$, $0.4$, $0.6$, and $0.8$ times the maximum value of the model, where the lowest level corresponds to the $3\sigma$ level of the emission.}
}
\label{fig:model_grid}
\end{figure*}

To derive the relative column density at each pixel with a size of $\rm 0''.04$, we summed over all surface densities $\Sigma \propto r_{\rm{cen}}^{-1.5}$ along the line of sight at each position on the plane of sky and averaged all column densities at the positions enclosed by the pixel. If we assume that the emission is optically thin, then the intensities would be proportional to the column densities. We performed Gaussian convolution on the simulated three-dimensional dataset to match the $\rm C^{18}O$ synthesized beam with dimensions of $\rm 0''.31 \times 0''.28$ ($\rm P.A.=-81^{\circ}$). Then, we extracted the simulated PV diagrams along {the axis at P.A.$=60^{\circ}$}. Further details regarding the model derivation can be found in Appendix \ref{sec:pv_derivation}.

In the PV diagram (Figure \ref{fig:C18OPV}(a)), the largest velocity offsets appear at ${\pm}\sim$0\farcs1 from the central source.
Due to the limited angular resolution of $\sim$0\farcs3, the radius of the centrifugal barrier is difficult to constrain.
We therefore constructed the model PV diagrams with different $r_{\rm CB}$ and central stellar masses, and compared with the observations. 
The angle of the disk normal is assumed to be 35$^{\circ}$ from the plane of the sky, which is between the axes of the outflow ($\sim$40$^{\circ}$, see Section \ref{subsubsec:outflow_modeling}) and the jet ($\sim$30$^{\circ}$, see Section \ref{subsubsec:jet_modeling}).
The results are shown in Figure {\ref{fig:model_grid}}.
The models with small $r_{\rm CB}$, such as 10 and 20 au, reproduce the triangle structures of the blue and red sides, while those with $r_{\rm CB} =$ 60 au, the emission in the first and third quadrants disappears. 
Models with $R_{\rm cb} = 10-40~\rm au$ and $M_{\star} = 0.08-0.1~M_{\odot}$ can reasonably reproduce the observed PV diagram.

\subsection{Outflow Configuration}
\subsubsection{CO Outflow Modeling}
\label{subsubsec:outflow_modeling}
\begin{figure*}
\epsscale{1.2}
\plotone{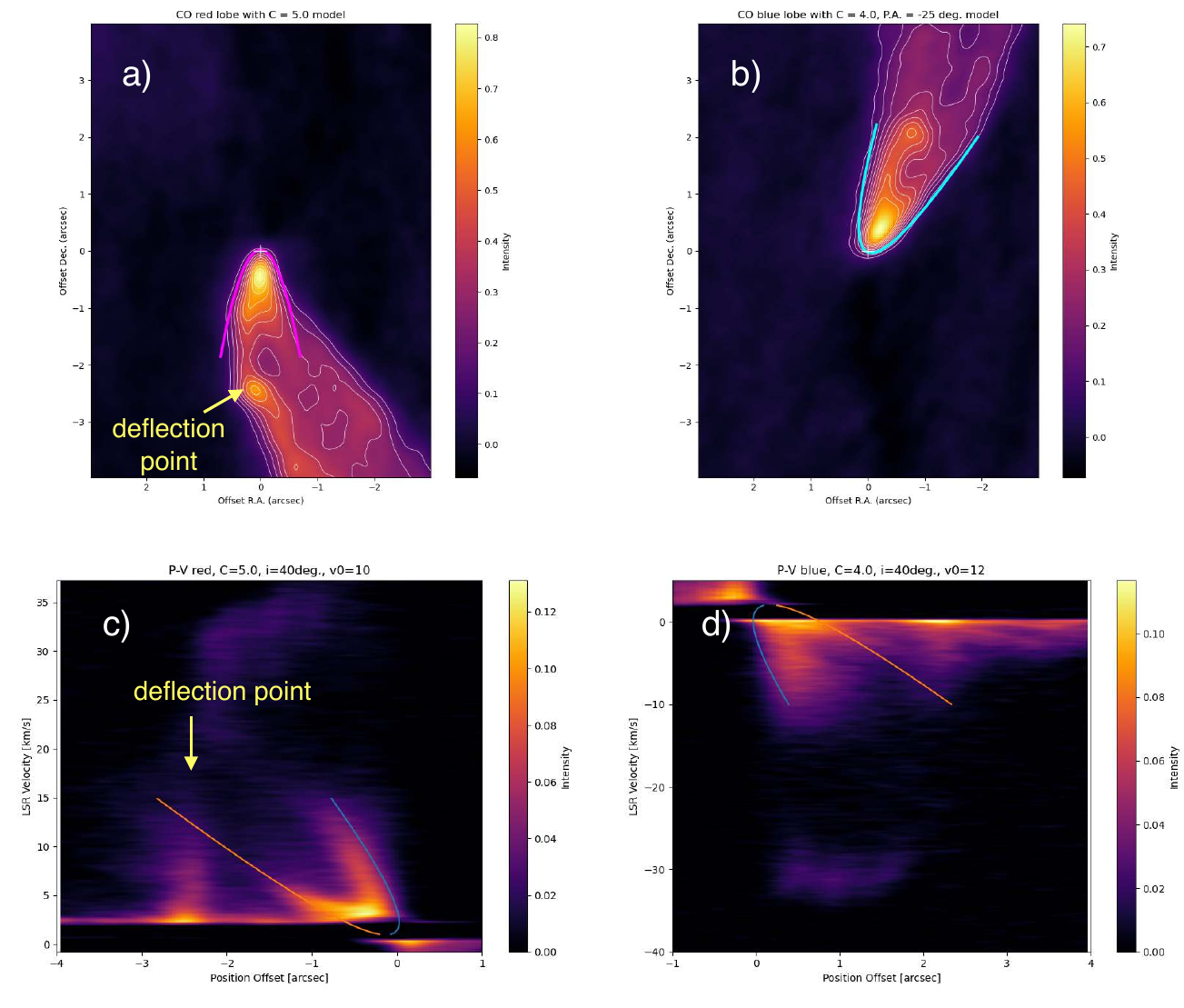}
\caption{{(a) Outflow shell model curve with $C =$ 5.0 arcsec$^{-1}$ (magenta curve) overlaid on the CO moment 0 map of the redshifted component.
Contours are drawn every 5~$\sigma$ (1~$\sigma$ = 15 mJy beam$^{-1}$ km s$^{-1}$) with the lowest contour level at 10~$\sigma$.
(b) Model curve with $C =$ 4.0 arcsec$^{-1}$ (cyan curve) overlaid on the CO moment 0 of the blueshifted component. 
Contour stars from 10~$\sigma$ (1~$\sigma$ = 10 mJy beam$^{-1}$ km s$^{-1}$) with a step of 5~$\sigma$.
(c) and (d) PV diagrams along the axes of the redshifted lobe, $\rm P.A. = 0^{\circ}$. (c), and the blueshifted lobe, $\rm P.A. = -25^{\circ}$. (d). 
Blue and orange curves are the model curves produced by $C =$ 5.0 arcsec$^{-1}$ and $i {\sim}$ 40$^{\circ}$ for (c), and $C =$ 4.0 arcsec$^{-1}$ and $i {\sim}$ 40$^{\circ}$ for (d).
{ The deflection point of the eastern wall of the southern lobe is indecated in panels (a) and (c).}}
\label{fig:outflow_model}}
\end{figure*}

Following \cite{Lee2000} and \cite{Jhan2022}, we used the wide-angle wind-driven shell model, which assumes that the outflow shell is parabolic and radially expanding, to simultaneously fit the $\rm CO$ outflow PV diagram and its outer spatial boundaries in the moment 0 map. In this model, the structure and velocity { of the} outflow shell can be described by the equations in cylindrical coordinates, $(z, R)$:
\begin{equation}
z = CR^2 
\end{equation}
\begin{equation}
v_R = v_0R
\end{equation}
\begin{equation}
 v_z = v_0z
\end{equation}
where {$C$ and $v_0$ are free parameters that  describe the curvature of the parabola near the base of the outflow lobe and velocity distributions of the outflow shell, respectively.}
 
{The curvature $C$ is determined so that the model curve reproduce the 10 $\sigma$ contour of the moment 0 maps.
The shape of the redshifted lobe near the base is reproduced with $C=$5.0 with a position angle of $\sim$0$^{\circ}$ (Figure \ref{fig:outflow_model}(a)), which is aligned with that of the jet.
Since the redshifted lobe is deflected to the southwest at the larger offset, we apply this model to the inner region with a projected distance from the protostar to be within $\pm$ $\sim$~2\arcsec.
The shape of the blueshifted lobe is reproduced with $C =$4.0 with a position angle of $\sim$~$-$25$^{\circ}$ (Figure \ref{fig:outflow_model}(b)), which is in the middle of those of the western wall ($\sim$~$-$22$^{\circ}$) and the EHV jet ($\sim$~$-$30$^{\circ}$).
}

{The inclination angles of the outflow axes were determined from the PV diagrams.
The PV diagram of the redshifted outflow exhibits clear parabolic shape near the base.
This parabolic shape can be reasonably reproduced by the model with $C=$5.0, $i=$40$^{\circ}$, and $v_0 =$~10 km s$^{-1}$ arcsec$^{-1}$.
The vertical emission spur seen at ${ {\sim}~{-}}$2\farcs5 { labeled in Figure \ref{fig:outflow_model}(c)} to the deflection point of the eastern wall.
The PV of the blueshifted lobe is reproduced by the model with $C=$4.0, $i=$40$^{\circ}$, and $v_0 =$~12 km s$^{-1}$ arcsec$^{-1}$, although the parabolic pattern in this lobe is vague.
}

{It should be noted that the inclination angles of the outflow axes derived here are the ones at the bases of the lobe{ s}. 
Since the CO emission at large distances mainly appears in the low velocity ranges, ${\lesssim}~{\pm}~6$ km s$^{-1}$ (see Figure \ref{fig:CO_channel_map}), the inclination angles of the larger scale outflow lobes could be smaller than those at the bases.}

\subsubsection{Physical Parameters of the outflow components}
\label{subsubsec:outflow}

Using the procedure outlined in \cite{Yildiz2015}, we derived physical parameters such as the total { mass,} momentum, outflow force, kinetic energy, and mechanical { power}. Following \cite{Yildiz2015} and \cite{Dutta2022}, we assume an excitation temperature of $T_{\rm ex} = \rm 50~K$ for the CO outflow and a { fractional} abundance of { CO to be} $X_{\rm CO} \sim 10^{-4}$. We also assume an LTE condition and that the CO emission is optically thin. 
{The inclination angles of the outflow axes were assumed to be { 40$^{\circ}$}.}
We summarize the derived physical parameters of the outflows in Table \ref{tab:outflow_measurements}. 

\begin{deluxetable*}{cccc} \label{tab:outflow_measurements}
\tablewidth{\textwidth}
\tablecaption{Quantities measured from the $\rm CO$ outflows.}
\tablehead{\colhead{Quantity} & \colhead{Blueshifted Lobe} & \colhead{Redshifted Lobe}}
\startdata
Spatial extension ($\rm au$) & $4140$ & $4140$  \\
Maximum velocity offset ($\rm km~s^{-1}$) & $16.6$ & $20.8$ \\
Dynamical timescale ($\rm yr$) & {990$^a$} & {790$^a$} \\
Momentum ($10^{-3}~M_{\odot}~\rm km~s^{-1}$) & ${2.39^{a}}$ & ${5.02^{a}}$ \\
Outflow force ($10^{-6}~M_{\odot}~\rm km~s^{-1}~yr^{-1}$) & ${2.4^{a}}$ & ${6.2^{a}}$ \\
Mass ($10^{-4} M_{\odot}$) & $5.53$ & $9.25$ \\
Kinetic energy ($10^{41} \rm erg$) & ${1.61^{a}}$ & ${4.58^{a}}$ \\
Mechanical { power} ($10^{-3} L_{\odot}$) & ${1.4^{a}}$ & ${4.8^{a}}$ \\
\enddata
\tablenotetext{a}{After the correction of the inclination angle.}
\end{deluxetable*}

\subsection{High-velocity Molecular Jets}
\subsubsection{SiO Jets and the Shock-forming Model}
\label{subsubsec:jet_modeling}
\begin{figure*}[ht!]
\epsscale{1.0}
\plotone{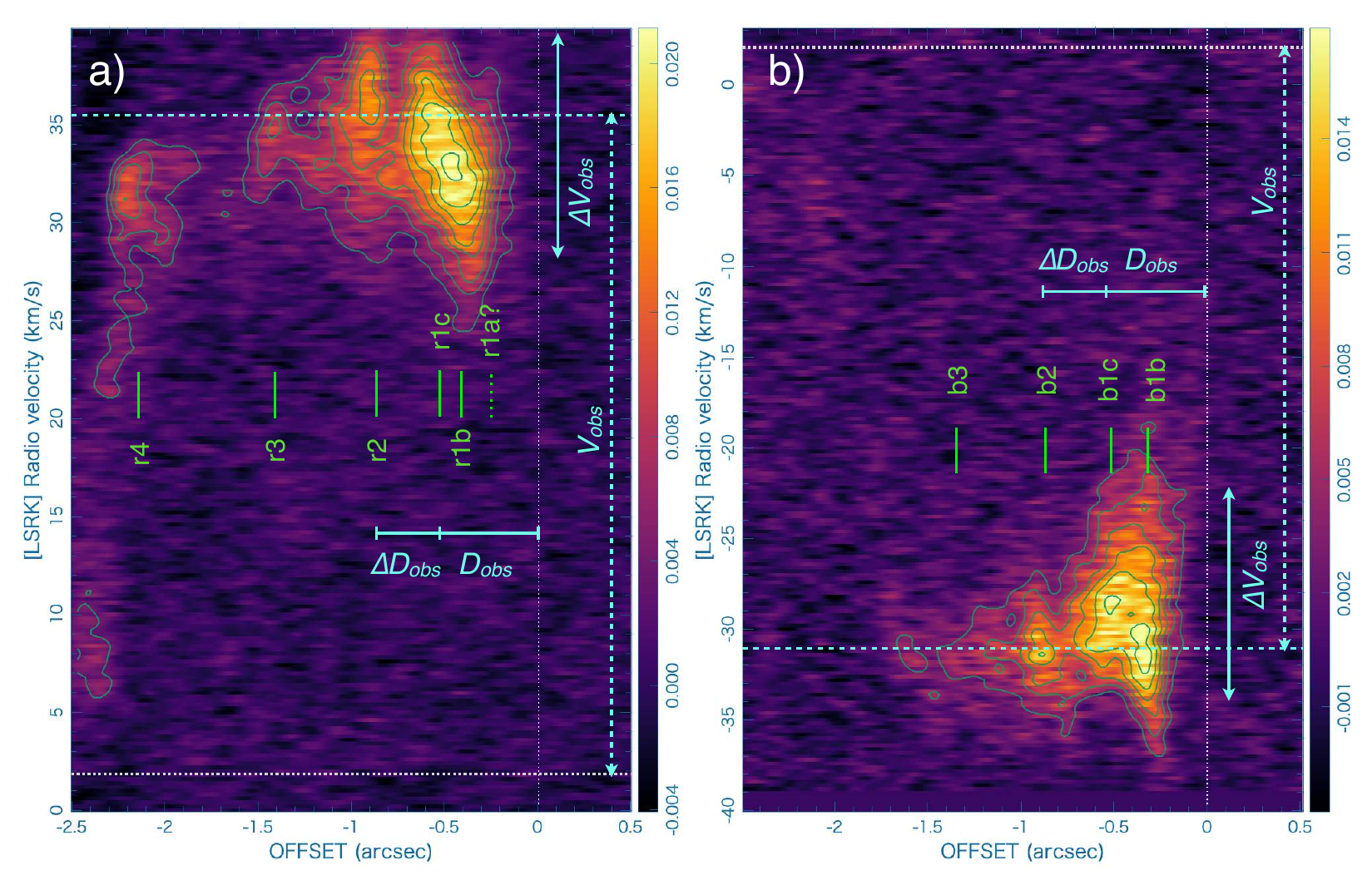}
\caption{{High resolution (TM1) SIO PV diagrams along the redshifted (a) and blueshifted (b) jet axes.
Contours are drawn every 2.4 mJy beam$^{-1}$ (2 $\sigma$) with the lowest contour level of 3.6 mJy beam$^{-1}$ (3 $\sigma$).
The locations of the knots are marked in green lines.
Annotated cyan arrows (dashed and solid) indicate the observed jet velocities $V_{\rm obs}$ and velocity dispersions ${\Delta}V_{\rm obs}$.
$D_{\rm obs}$ and $\Delta D_{\rm obs}$, used for the jet modeling are indicated by cyan horizontal lines.}
 Note that velocity offsets larger than ${\pm}$40 km s$^{-1}$ were not covered by the present correlator setting.
 \label{fig:SiO_jet}}
\end{figure*}

To derive physical parameters of the $\rm SiO$ jets, we adopt the shock-forming model proposed by \cite{Jhan2022}, which assumes that the internal shocks (so-called “knots”) were formed by the fastest material catching up and colliding with the slowest material due to periodic variations in the velocities of the jet ejections. We use the model to derive the jet inclination angle $i$, the mean jet velocity $V_j$, and the period of the variation $P$. These can be related to observed quantities in the following equations:
\begin{equation}
i = \tan^{-1}(\frac{V_{\rm{obs}}}{\pi \frac{D_{\rm{obs}}}{\Delta D_{\rm{obs}}} \Delta V_{\rm{obs}}}) 
\end{equation}
\begin{equation}
V_j = \frac{V_{\rm obs}}{\sin {i}}
\end{equation}
\begin{equation}
P = \frac{\Delta D_{\rm{obs}}}{V_j \cos i}
\end{equation}
where $V_{\rm{obs}}$ is the observed (projected) mean jet velocity, $\Delta V_{\rm{obs}}$ is the observed velocity variation, $D_{\rm{obs}}$ is the observed distance between the first knot and the central protostar, and $\Delta D_{\rm{obs}}$ is the observed distance between the first and second knots. 

Figures \ref{fig:SiO_jet}(a) and {(b)} show the SiO PV diagrams cut along the major axes of the {redshifted} and {blueshifted} jets, respectively. 
The TM1-only dataset was used in order to spatially resolve the knots. 
{In the PV diagram, the three subknots in r1 are not clearly separated; especially the presence of the r1a knot is not clear.
On the other hand, the r1c is identified as a secondary peak at $V_{\rm LSR}{\sim}$~35 km s$^{-1}$ with emission { extending} to $>$~40 km s$^{-1}$.
Within the innermost knot pair, r1b and b1b, an increase in velocity is observed with increasing distance from the central star.
Such an apparent acceleration, which is also observed at the base of the L1448C(N) jet \citep{Hirano2010,Yoshida2021}, is different from that expected in the sideways ejection of the shocked region.
In the case of the L1448C (N) jet, \citet{Jhan2022} argued that the first pair of knots is unlikely to be formed by shocks, because the velocity dispersion of $>$40 km s$^{-1}$ observed in these knots is too large for the SiO to survive \citep[e.g.][]{Schilke1997,Gusdorf2008}. 
Although the velocity dispersions in r1b and b1b, $\sim$20 km s$^{-1}$, being within the range of SiO to survive, we excluded r1b and b1b from our analysis, and assumed that the first knot pair produced by shock are r1c and b1c.}
By visually inspecting the PV diagrams, {we estimated $D_{\rm obs}$ and $\Delta D_{\rm obs}$ to be $0''.54 \pm 0''.1$ and $0''.36 \pm 0''.1$, respectively,} for both jet components, where the uncertainties were assumed to be the geometric mean of the TM1 beam dimensions. 

{The radial velocity of the jet, as referenced in $V_{\rm obs}$, was determined by performing Gaussian fits on the representative spectral profiles at the specified knot positions of r2 and b2. The spectra corresponding to r1c and b1c were omitted due to their deviation from Gaussian spectral shapes. 
The radial sideways-ejection velocity, indicated as ${\Delta}V_{\rm obs}$, was evaluated using the line width measured at a 3 $\sigma$ level of the line profiles at the knots for r2 and b1c.
}
{We derived $V_{\rm obs} \sim +34.0 \pm 0.1\rm ~km~s^{-1}$ and $\Delta V_{\rm obs} \sim 12.0\rm ~km~s^{-1}$ for the redshifted jet and $V_{\rm obs} \sim -32.0 \pm 0.06\rm ~km~s^{-1}$ and $\Delta V_{\rm obs} \sim 11.5 \rm ~km~s^{-1}$ for the blueshifted jet. 
}
From these measurements, {we derived the inclination angle, mean deprojected jet velocity, and period of velocity variation to be  $i \sim 31 \pm 12^{\circ}$, $V_{\rm j} \sim 66 \pm 23\rm ~km~s^{-1}$, and $P \sim 11 \pm 8~\rm yr$ for the redshifted jet and $i \sim 31 \pm 12^{\circ}$, $V_{\rm j} \sim 63 \pm 22\rm ~km~s^{-1}$, and $P \sim 11 \pm 9~\rm yr$ for the blueshifted jet}. 
{It should be noted that the current correlator configuration does not fully cover the velocity range of the redshifted jet. 
Therefore, the inclination angle of the redshifted jet could be smaller if the full velocity dispersion is measured.
Assuming that the line profile at r2 knot is symmetric to its peak velocity, $\Delta V_{\rm obs}$ could be $\sim$14.8 km s$^{-1}$.
In this case, the inclination angle is estimated to be $\sim$26$^{\circ}$.}

\subsubsection{The Mass Loss Rate and $CO/SiO$ Abundance} 
The CO column density in the jet was calculated using the Equation (\ref{eqn:optically_thin}) assuming the LTE condition and optically thin line emission.
The CO excitation temperature in the jet was assumed to be $\sim$150 K following \cite{Dutta2022}.
The moment 0 maps used were integrated over the velocity ranges of $-37 < v_{\rm{obs}} < -20$ km s$^{-1}$ and $23.4 < v_{\rm{obs}} < 38$ km s$^{-1}$. 
The CO column densities averaged over the blueshifted and redshifted jets are $\sim 2.5\times10^{16}$ and $\sim 2.7\times10^{16} \rm ~cm^{-2}$, respectively.
A mass-loss rate can be estimated from the beam-averaged $\rm CO$ column densities by assuming a constant abundance ratio ($X_{\rm{CO}} = N_{\rm{CO}}/N_{\rm{H_2}}$), as shown in the following equation:
\begin{equation}
\label{eqn:mass_loss_rate}
\dot{M_j} = \mu_{\rm{H_2}} m_{\rm{H}} \frac{N_{\rm{CO}}}{X_{\rm{CO}}} V_j b_m
\end{equation}
where $\mu_{\rm{H_2}}=2.8$ is the mean molecular weight, $m_{\rm{H}} = 1.67\times 10^{-24}~\rm g$ is the mass of a hydrogen atom, $V_j$ is the mean deprojected jet velocity derived in Section \ref{subsubsec:jet_modeling}.
{Since the transverse width of the jet, $\sim$27 au, is not resolved by the beam of the CO image (Figure \ref{fig:jet_mom0}a), we adopted the geometrical mean of the CO beam, $b_m = \sqrt{0''.31 \times 0''.28} \sim 0''.3$ ($\rm 108~au$), to be the jet width. }
We assumed $X_{\rm{CO}} \sim 4 \times 10^{-4}$ \citep{Glassgold1991}, regarding this to be the upper limit as lower CO abundances have been derived in some cases (e.g., \citealp{Hirano2010}). 
From Equation (\ref{eqn:mass_loss_rate}), we derived the  jet mass-loss rates to be  ${\sim (4.5 \pm 1.6) \times 10^{-8} ~M_{\odot}~\rm yr^{-1}}$ and ${\sim (4.7 \pm 1.6) \times 10^{-8} ~M_{\odot}~\rm yr^{-1}}$ for the blueshifted and redshifted components, respectively. 
The derived total mass-loss rate is thus ${\sim (9.2 \pm 3.2) \times 10^{-8} \rm ~M_{\odot}~\rm yr^{-1}}$. 
The mechanical powers of the blue- and redshifted jets defined as $L_{\rm j} = (1/2){\times}\dot{M_j}{V_j}$ are $\sim${1.5}$\times$10$^{-2}L_{\odot}$ and $\sim${1.7} $\times$10$^{-2}L_{\odot}$, respectively.
The total mechanical power of the jet is $\sim$ {3.1}$\times$10$^{-2} L_{\odot}$, which is {$\sim$ 5 times larger than that of the outflow, $\sim${6.2}$\times$10$^{-3} L_{\odot}$}.

Similar to the $\rm CO$ analysis, we derived the mean $\rm SiO$ column densities using Equation (\ref{eqn:optically_thin}) to be $\sim 2.5 \times 10^{14}$ and $\sim 1.8 \times 10^{14} \rm ~cm^{-2}$ for the blueshifted and redshifted jets, respectively, from the TM1+TM2 data, under the same assumptions of optical thin emission and LTE with an excitation temperature of $\rm 150~K$. The  $\rm SiO/CO$ abundance values are $\sim 4.3 \times 10^{-3}$ for the blueshifted jet and $\sim 5.8 \times 10^{-3}$ for the redshifted jet. 
The SiO/CO abundance in the first redshifted knot should be higher than the estimated value because the CO emission is barely seen in this knot despite the bright SiO emission from the knot.

\subsection{Summary of the derived physical parameters} \label{subsec:parameters}

In Table \ref{tab:summary}, we summarize the derived physical parameters of the central source and envelope/disk, and position angle, inclination angle, and velocity of the outflow and jet.

\begin{table*}[htbp]
\centering
\caption{Summary of Derived Physical Parameters of G204NE}
\label{tab:summary}
\begin{tabular}{lccc}
\hline\hline
Component & & Parameter & Value \\
\hline
Central Source & & $L_{\rm bol}$ & 1.15 $L_{\odot}$ \\
 & & $T_{\rm bol}$ & 33 K \\
 & & $M_{*}$ & 0.08--0.1 $M_{\odot}$ \\
\hline
Envelope/Disk & Extended component & FWHM& 780$\pm$26 au \\
 & & Mass & $\gtrsim$ 0.13 $M_{\odot}$ \\
 & Intermediate component & FWHM & 83$\pm$3.6 au  \\
 & & Mass & $\gtrsim$ 0.049 $M_{\odot}$ \\
 & Compact component (Disk) & FWHM & 24$\pm$0.76 au \\
 & & Mass & $\gtrsim$ 0.015 $M_{\odot}$ \\
\hline
Outflow  & Northern lobe(blue)& P.A. & $-$5$^{\circ}$ (NE), $-$20$^{\circ}$ (NW) \\
 & & $i_{\rm flow}$ & $\sim$40$^{\circ}$\\
 & & $V_{\rm flow}^{~~~a}$ & 25.8 km s$^{-1}$\\
 & Southern lobe (red) & P.A. & 190$^{\circ}$ (SE), 205$^{\circ}$ (SW) \\
  & & $i_{\rm flow}$ & $\sim$40$^{\circ}$\\
 & & $V_{\rm flow}^{~~~a}$ & 32.4 km s$^{-1}$\\
\hline
Jet  & Northern jet (blue) & P.A.  & $-$30$^{\circ}$ \\
 & & $i_{\rm jet}$ & $\sim$31$\pm$12$^{\circ}$ \\
  & & $V_{\rm jet}$ & 63$\pm$22 km s$^{-1}$ \\
 & Southern jet (red) & P.A. & 90$^{\circ}$ \\
 & & $i_{\rm jet}$ &  $\sim$31$\pm$12$^{\circ}$\\
  & & $V_{\rm jet}$ & 66$\pm$23 km s$^{-1}$ \\
\hline
\end{tabular}
\tablenotetext{a}{After the correction of the inclination angle.}
\end{table*}

\section{Discussion} \label{sec:discussion}

\subsection{Evolutionary Phase of the Central Source}
The highly collimated outflow with a deprojected velocity of $\sim$~{30} km s$^{-1}$ and the extremely high-velocity (EHV) jet with $V_j~{\sim}~${65} km s$^{-1}$ indicate that the central source of G204NE has already reached the protostellar stage.
Meanwhile, the continuum emission from the central source is generally faint at wavelengths shorter than $\rm 70~\mu m$ and exhibits a low bolometric temperature of $T_{\rm{bol}} \sim 33\rm ~K$.
The SED of this source with a peak around 160 $\mu$m resembles those of the first hydrostatic core candidates compiled by \citet{Dutta2022}, implying that the central star is in the early stage of Class 0 as in the case of G208Walma \citep{Dutta2022}.
Our analysis of the continuum visibility profile unveiled the presence of a spatially compact circumstellar disk with a radius of $\sim 12\rm ~au$, {which also corroborates the scenario that G204NE is in the nascent} stage of the Class 0 phase.
The collimated outflows with cavity angles of $\sim {40^{\circ}}$ {at their bases} are also characteristic of an early Class 0 source (e.g., \citealp{Aso2017, Aso2018, Dutta2020}). 
Furthermore, the chemical properties of G204NE also support this interpretation. The C$^{18}$O emitting region is spatially compact with a radius of $\sim$240 au, and is surrounded by the spatially extended emission of the deuterated molecules. 
This indicates that the central protostar has just started heating its surrounding, and that the bulk of the gas in the core remains colder than the CO {sublimation} temperature.
The $\rm C^{18}O$ abundance toward the protostellar position is $\lesssim 10^{-8}$, which is one order of magnitude lower than that of the interstellar value.
This is in part because the bulk of the envelope gas along the line of sight is colder than {the $\rm CO$ sublimation temperature of $\sim 25\rm ~K$}, except for the close vicinity of the central source.
Another possibility is the grain surface chemistry.
During the prestellar phase, CO on the grain surface decreases due to the conversion into more complex molecules such as CH$_3$OH (e.g., \citealp{AlonsoAlbi2010, Fuente2012}).
Meanwhile, the drop in $\rm N_2D^+$ abundance toward the center suggest the evaporation of $\rm CO$ from the dust and the reaction between $\rm N_2D^+$ and the $\rm CO$ gas to form $\rm DCO^+$.

\subsection{Mass Accretion and Ejection}
The mass accretion rate $\dot{M}_{\rm acc}$ can be estimated using the bolometric luminosity $L_{\rm bol}$ and the protostellar mass $M_*$:
\begin{equation}
    \label{eqn:Macc}
    {\dot{M}_{\rm acc} {\sim} \frac{L_{\rm bol}R_*}{GM_*} M_{\odot}{\rm yr}^{-1}.}
    \label{eqn:mass_acc_rate}
\end{equation}
Here, we assume that the bolometric luminosity of the central protostar $\sim${1.15} $L_{\odot}$ originates entirely from mass accretion.
The mechanical luminosities of the jet and outflow, {0.03} $L_{\odot}$ and {0.006} $L_{\odot}$, respectively, are not taken into account because they are negligible as compared to the bolometric luminosity.
The mass of the central star $M_*$ derived in Section \ref{subsec: kinematics} from the envelope kinematics is {$\sim$ 0.1} $M_{\odot}$.
The stellar radius $R_*$ is assumed to be $\sim$2$R_{\odot}$ \citep{Stahler1988}.
The mass accretion rate derived from Equation (\ref{eqn:mass_acc_rate}) is $\dot{M}_{\rm acc}{\sim} {7.3}{\times}10^{{-7}} M_{\odot}$ yr$^{-1}$.

The ratio of $\dot{M}_{\rm jet}/\dot{M}_{\rm acc}$ becomes $\sim${0.13}.
This value is consistent with the $\dot{M}_{\rm jet}/\dot{M}_{\rm acc}$ ratio for the Class 0/I jets in the literature (\citealp{Lee2000, Ellerbroek2013}).
The $\dot{M}_{\rm jet}/\dot{M}_{\rm acc}$ ratio enable us to estimate the footpoint radius of the jet following the method described in \cite{Lee2020}. 
The $\dot{M}_{\rm jet}/\dot{M}_{\rm acc}$ ratio is related to the magnetic lever arm parameter $\lambda_{\phi}$ (e.g., \citealp{Pudritz2007}),  
\begin{equation}
    \lambda_{\phi}\,{\equiv}\,\left ( \frac{r_{\rm A}}{r_0}\right)^2\,{\approx}\,\frac{\dot{M}_{\rm acc}}{\dot{M}_{\rm jet}},
    \label{eqn:magnetic lever arm}
\end{equation}
where $r_0$ is the footpoint radius of the jet on the disk and $r_{\rm A}$ is the Alfv\'{e}n radius along the streamline starting from $r_0$.
In the case of the magneto-centifugally launched jet/outflow, the terminal velocity of the jet/outflow $v_{\rm jet}$ and the Keplerian velocity at the footpoint has a relation,
\begin{equation}
    v_{\rm jet}\,{\sim}\,(2{\lambda}_{\phi}-3)^{1/2} v_{\rm kep},
\end{equation}
where $v_{\rm kep}$ is the Keplerian velocity at the footpoint $r_0$, i.e. $(GM_*/r_0)^{1/2}$.
Thus, the footpoint radius can be estimate by
\begin{equation}
    r_0\,=\,(2\lambda_{\phi}-3)\frac{GM_*}{v_{\rm jet}^2}.
\label{eqn:footpoint_radius}
\end{equation}
In the case of G204NE, the mass of the central star, the mean velocity of the jet, and the magnetic lever arm parameters are $\sim${0.1} $M_{\odot}$, $\sim${65} km s$^{-1}$, and $\sim${8.0}, respectively.
Using these parameters, the footpoint radius calculated from Equation (\ref{eqn:footpoint_radius}) is $\sim${ 0.28} au.

The footpoint radius of the jet derived here is {within twice of} the dust sublimation radius of $R_{\rm sub}\,{\sim}\,0.15{\times}\sqrt{L_{\rm bol}/L_{\odot}}$ \citep{Yvart2016} $\sim$0.16 au.
This implies that the jet could be launched from the outer edge of the dust sublimation zone.
A comparison of the measured SiO/CO abundance ratio and jet mass loss rate, $\sim$5${\times}$10$^{-3}$ and $\sim$1.2$\times$10$^{-7}$ $M_{\odot}$, respectively, with the chemical models of the jet with different dust-to-gas mass ratios by \citet[their Figure 12]{Tabone2020} suggests that the jet contains small amount of dust (a few percent of the canonical ISM value).

\subsection{Origin of Multiple Outflow Components}
As shown in Figure \ref{fig:outflow_mom0}, the tips of the northern and southern lobes exhibit a pair of bow-like patterns, implying that each lobe consists of two lobes with slightly different position angles.
{ As summarized in Table \ref{tab:summary}, the} axes of the two lobes are ${\sim}-$5$^{\circ}$ (NE lobe) and ${\sim}-$20$^{\circ}$ (NW lobe) in the northern lobe, and ${\sim}$190$^{\circ}$ (SE lobe) and ${\sim}$205$^{\circ}$ (SW lobe) in the southern lobe.
The velocity asymmetry observed in the northern lobe—i.e., the lower velocity in the eastern wall and the higher velocity in the western wall—is naturally explained if the NE and NW lobes have different inclination angles; the NE lobe having a smaller inclination angle with respect to the plane of the sky contributes to the emission of the eastern wall, and the NW lobe with larger inclination angle contributes to the emission of the western wall.
In the southern lobe, the axis of the SW lobe is close to the plane of the sky, and that of the SE lobe is more inclined.
The SW lobe is likely the redshifted counterpart of the NE lobe, while the SE lobe is likely the counterpart of the NW lobe.
Outflow systems with multiple lobe pairs with different orientations are also observed in the early Class 0 protostars such as HOPS 373SW \citep{Lee2024}.

A possible explanation for the two pairs of lobes is an embedded system of close binary protostars, each of which is driving a bipolar outflow.
Although the central source of G204NE does not show a signature of binary down to the scale of $\sim${ 18} au, embedded binary with a separation smaller than { this scale}  cannot be excluded.
A three-dimensional magnetohydromagnetic (MHD) simulation by \citet{SaikiMachida2020} has revealed that the close binary system with a separation of $<$ 30 au can be formed by means of fragmentation of the collapsing core.
Each protostar in this binary system is launching a highly-collimated outflow/jet.
Since the axes of the outflows/jets are distorted due to the orbital motion of the launching points, the misaligned outflows/jets can be naturally formed.
In this scenario, the EHV jet observed in CO, SO, and SiO is driven by one of the binary protostars, the mass accretion toward which is currently active.
As discussed in \citet{SaikiMachida2020}, the twin outflows/jets can be detectable only in the region close to the protostars where the outflows/jets are not significantly disturbed by the orbital motion of the binary.
In the outer region, the two outflows/jets are difficult to recognize because the outflows/jets are highly tangled.
In the case of the binary system with a separation of $a$ and the total mass of $M_{\rm tot}$, the orbital period $P$ is given by
\begin{equation}
    P\,=\,(4{\pi}^2a^3\,/\,GM_{\rm tot})^{1/2}.
    \label{eq:binary_period}
\end{equation}
For the putative binary system having a separation of $\sim${ 18} au and a total mass of ${\sim}${0.1} $M_{\odot}$, the orbital period $P$ is calculated to be $\sim$ { 240 yr}.
The orbital period is even shorter for the binary with a smaller separation.
Assuming that the maximum radial velocity of the outflow is $\sim$20 km s$^{-1}$ and the inclination angle from the plane of the sky is $\sim${ 40}$^{\circ}$, the outflow can reach $\sim${ 1200 au} in one orbital period.
Thus, the twin outflows/jets could be detectable only in the region much smaller than $\sim${ 1200 au}. 
This implies that the observed twin bow-shock structures located at $\sim$4100 au from the central source are unlikely to be attributed to the twin outflows/jets from the close binary protostars.
 
Another possible scenario is that the two pairs of lobes and an EHV jet are driven by a single protostar. 
In the turbulent core, the angular momentum distribution is non-uniform. 
The orientation of the disk formed in such an environment is expected to be time-variable due to the accretion of the gas with nonuniform angular momentum.
This phenomenon is particularly pronounced in the youngest protostars such as G204NE, whose disk mass is notably small.
{Assuming that the S-shaped feature observed in dust continuum (Figure \ref{fig:moment0_dense}) is feeding the material to the disk, the reorientation of the angular momentum of the disk can be estimated following the method of \citet{Goddi2020}.}
{The total angular momentum of the Keplerian rotating disk around the central star with $M_{\star}$is estimated as
\begin{equation}
    l_{\rm disk} = 0.4~M_{\rm disk}~\sqrt{GM_{\star}R_{\rm disk}}~,
\end{equation}
where $M_{\rm disk}$ and $R_{\rm disk}$ are the mass and radius of the disk.
The total angular momentum of the disk  with a radius of 12 au and a mass of 0.015 $M_{\sun}$ rotating around the central star of $\sim$0.1 $M_{\odot}$ is calculated to be $l_{\rm disk}~{\sim}$~5.8 $\times$ 10$^{50}$ g cm$^2$ s$^{-1}$.
If half of the mass of the S-shaped feature, $\sim$0.005 $M_{\odot}$ impact the disk at $r =$ 12 au with a free-fall velocity, ${ v_{\rm inf\_vertical}}$ = 3.8 km s$^{-1}$, the angular momentum of the falling gas is $l_{\rm dump} {\sim}$~5.3 $\times$ 10$^{50}$ sin $\psi$ g cm$^2$ s$^{-1}$.
Here, $\psi$ represents the angle between the gas trajectory and the plane of the disk.
Since the C$^{18}$O emission is blueshifted toward the position of the northern arm and redshifted in the southern arm, the angle between the arms and disk, $\psi$, should be $<$55$^{\circ}$, because the disk is is inclined by $\sim$65$^{\circ}$ from the plane of the sky.
Assuming that $\psi$ is  $\sim$50$^{\circ}$, the angular momentum of the disk is reoriented by ${\theta} = {\rm tan}^{-1}~(l_{\rm dump}/l_{\rm disk})~{\sim}~ 42^{\circ}$
Given that the difference in position angle between the NE (SE) and NW (SW) lobes projected onto the plane of the sky is 15$^{\circ}$, the alteration in the disk's angular momentum is sufficient to account for the reorientation observed in the outflow directions.}

Moreover, the misalignment between angular momentum and the magnetic field is also of significant importance; three-dimensional magnetohydrodynamic (MHD) simulations conducted by \citet{HiranoMachida2019} and \citet{Machida2020} have demonstrated that if the initial angular momentum of the collapsing core is misaligned with the global magnetic field, the directions of the outflows fluctuate randomly over time.
Their models also show that the different velocity components of outflows/jets are ejected to the different directions due to the warped disk/pseudodisk structure.
This can also explains why the axes of the EHV jets are different from those of the larger scale outflow lobes.
The local misalignment between angular momentum and magnetic field occurs naturally in the turbulent core, in which angular momentum of the infalling material randomly changes with time and the magnetic field lines are inclined with respect to the initial direction \citep{Matsumoto2017}.

\subsection{Possible Mechanisms of outflow/jet deflection}
 
As shown in Figures \ref{fig:outflow_mom0} and \ref{fig:jet_mom0}, the orientations of the blue- and red-shifted components of the two pairs of outflows, as well as the jet, do not exhibit antiparallel alignment.
The axes of each three pairs are misaligned by $\sim$30$^{\circ}$.
Similar outflows with deflected lobes have been observed in several protostars (e.g., \citealp{Ching2016, Tobin2016, Aso2017, Yen2017, Aso2018}). 
Four possible mechanisms could explain the formation of deflected outflows:
(1) orbital motion of the binary protostar \citep{Fendt1998,MasciadriRaga2002},
(2) dynamical interaction with the dense gas \citep{Fendt1998,Umemoto1991},  (3)  
Lorentz forces between the external magnetic field 
and the current in the outflow lobe \citep{Fendt1998}, and (4) turbulent accretion in moderately magnetized core \citep{Matsumoto2017, Takaishi2024}.

The first mechanism involving binary orbital motion is unlikely, as the central star of G204NE lacks the signature of binary. 
It is also unlikely that the central star is a close binary system unresolved by the current resolution of $\sim${ 18} au, because the orbital motion of the close binary system produces a mirror symmetric wiggling pattern (e.g., \citealp{MasciadriRaga2002}) rather than inducing deflection.
In the second mechanism of dynamical interaction, as outflows propagate and encounter varying densities and gas compositions within the dense molecular cloud surrounding the central protostar, these inhomogeneities can cause the outflow trajectory to bend \citep{Fendt1998,Umemoto1991}. 
{A close-up view of the redshifted lobe (Figure \ref{fig:outflow_model}(a)) exhibits that the eastern wall of the redshifted lobe changes its direction at $\sim$2\farcs5 south of the central source where the bright emission spot is observed.
Since the dense ambient gas traced by N$_2$D$^+$ and DCO$^+$ extends to the southeast of the central source (dashed ellipse in Figure \ref{fig:moment0_dense}(c) and (e)), the interaction with this component could bend the direction of the flow to the southwest. 
The moment 1 image of DCO$^+$ (Figure \ref{fig:CO_DCOp_SiO}(a)) shows that the redshifted emission is extending from southeast to southwest of the southern lobe.
However, this redshifted DCO$^+$ emission does not extend to the { deflection} point; the ambient gas traced by the DCO$^+$ emission near the { deflection} point is blueshiftd rather than redshifted.
Therefore, the outflow { deflection} is unlikely to be attributed to the interaction with the dense ambient gas.
The observed { deflection} is likely to be due to the interaction between the jet and outflow.
As shown in Figure \ref{fig:CO_DCOp_SiO}(b), the { deflection} point corresponds to the location at which the jet impacts on the eastern wall of the redshifted lobe. }

\begin{figure*}
\epsscale{1.1}
\plotone{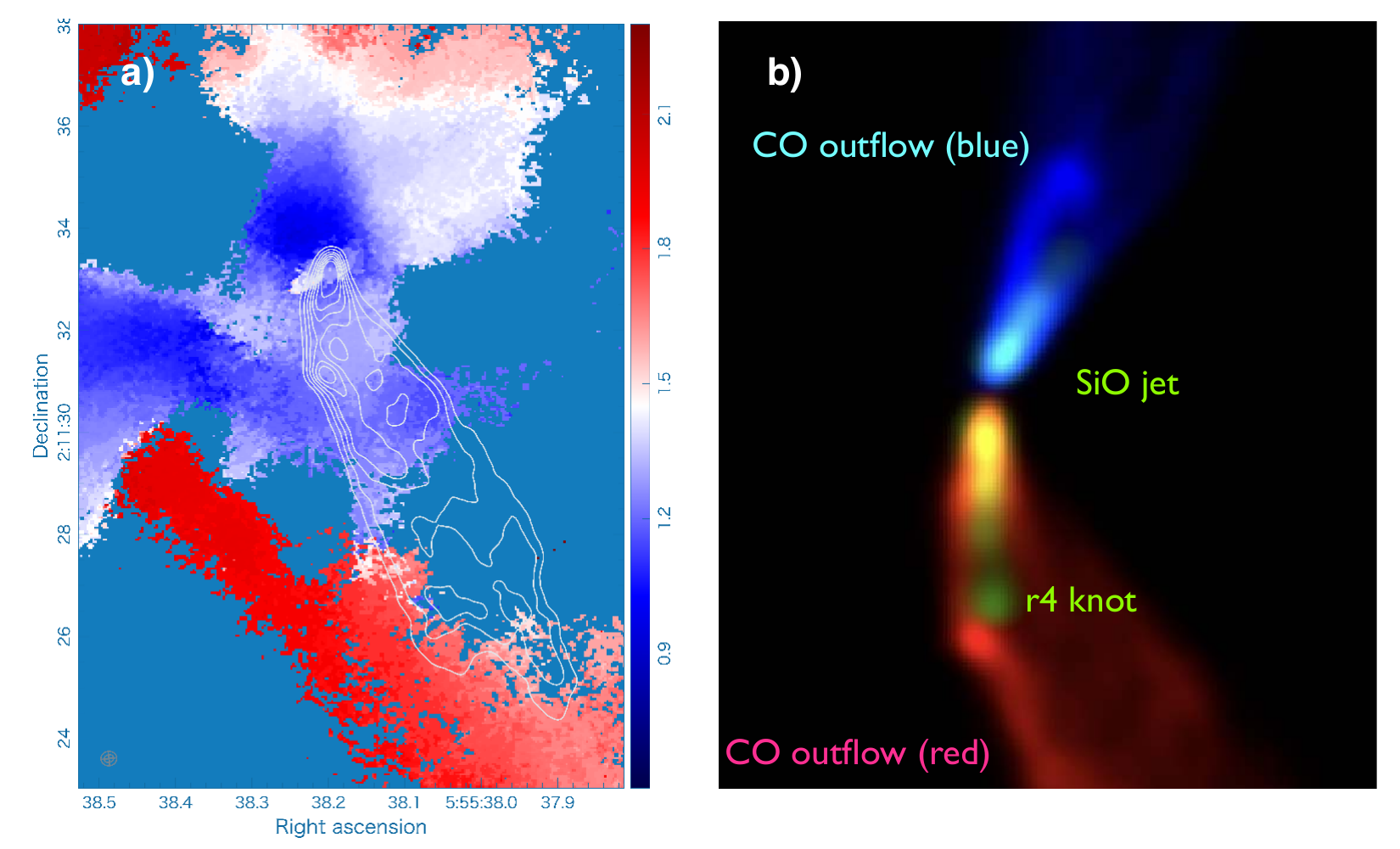}
\caption{{  (a) DCO$^+$ moment 1 image (color) overlaid on the redshifted CO outflow (contour).
Contours are drawn every 5 $\sigma$ (1 ${\sigma} =$15 mJy beam$^{-1}$ km s$^{-1}$) with the lowest contour level at 10 $\sigma$.
(b) CO outflow (blue and red) and SiO jet (green) at 0\farcs3 resolution. The location of the r4 knot is indicated. }
\label{fig:CO_DCOp_SiO}
}
\end{figure*}

The deflected structure can also be formed by the Lorentz force if the poloidal current in the jet/outflow and the counter jet/outflow flow the same direction \citep{Fendt1998}.
While the typical interstellar magnetic field of a few tens of microgauss is not strong enough to bend the jet/outflow, the magnetic field associated with one outflow functions as the external magnetic field for an adjacent outflow lobe.
This mechanism of electromagnetic interaction between adjacent lobes was examined to explain the similar deflected outflow in the protobinary system NGC1333 IRAS 4A \citep{Ching2016} and Serpens SMM4 \citep{Aso2018}.
Given that G204NE exhibits a pair of outflow lobes, it is plausible that the same mechanism could account for the deflection observed in this system. 
The strength of the magnetic field $B$ that is required to deflect the outflow/jet by ${ \phi}$ can be derived utilizing the subsequent equation:
\begin{equation}
B = {0.55\rm ~mG}~\frac{\sin  \phi}{\tan i_b} \frac{v'}{10~\rm km~s^{-1}} \left(\frac{L'}{1000\rm ~au}\right)^{-1}
\end{equation}
where $i_b$, $v'$, and $L'$ are the inclination angle, observed line-of-sight velocity, and projected length of the outflow lobe/jet, respectively \citep{Aso2018}. 
The deflection angle $\alpha$ corresponds to half of the misalignment of the axes, i.e. $\sim$15$^{\circ}$.
If we adopt (${ \phi}, i_b, v', L'$) to be (15$^{\circ}$, 25$^{\circ}$, 20 km s$^{-1}$, 4140 au), 
the required magnetic field strength would be $B \sim$ 150 $\mu$G. 
Stronger magnetic field is required to bend the jet with higher velocity and shorter length; adopting ($\phi, i_b, v', L'$) for the jet to be (15$^{\circ}$, 22$^{\circ}$, 35 km s$^{-1}$, 800 au), the required magnetic field strength is estimated to be $\sim$1.5 mG. 
Given that the magnetic field strength observed in typical protostellar envelopes is on the order of a few mG \citep{Girart2006, Hull2017, Aso2021}, it is plausible to quantitatively attribute the deflection mechanisms to the influence of the Lorentz force.

As observed with the outflow in Serpens SMM4B \citep{Aso2018}, the deflection of the outflow lobes and jet in G204NE appear in proximity to the central star.
\citet{Aso2018} suggested the possibility that the deflection of the outflow lobe might be attributed to disk-scale structures.
MHD simulations incorporating turbulence have demonstrated that misaligned outflow lobes can develop within a turbulent core where the magnetic field geometry is disturbed due to turbulence \citep{Matsumoto2017, Takaishi2024}.
The outflows propagate in the average direction of the magnetic field at that scale, resulting in the formation of an outflow lobe and counter-lobe that are not antiparallel.
The deflected outflow has been successfully reproduced in the MHD simulation of the moderately magnetized turbulent core \citep{Takaishi2024}.
In turbulent accretion models, the density distributions in the infalling envelope exhibit arc-like or spiral structures \citep{Matsumoto2017, Takaishi2024}, which is similar to the S-sped pattern seen in the TM1+TM2 continuum image (Figure \ref{fig:moment0_dense}).
Additionally, the turbulent accretion model is consistent with the interpretation of the multiple pairs of lobes and the jets exhibiting varying orientations, as discussed in the previous section.

\section{Conclusions} \label{sec:conclusions}
We have observed the molecular cloud core G204NE in the Orion B GMC with ALMA in Band 6. The physical and chemical characteristics of the structural components, the dense gas tracers, the outflow/jets, and the envelope kinematics have been investigated. Our main results are summarized below:
\begin{enumerate}
    \item The emission from G204NE is either barely detected or undetected at wavelengths $\lambda < \rm 70~\mu m$.  
    The bolometric temperature and luminosity derived from the SED analysis are $T_{\rm{bol}} \sim 33 \rm ~K$ and ${L_{\rm{bol}} \sim 1.15 \rm ~L_{\odot}}$, respectively, suggesting that this core harbor a low luminosity Class 0 protostar. 
    \item The continuum images of TM1 CLEAN $\sim$28 au resolution and SpM $\sim$18 au resolution reveal a compact central structure. 
    There is no clear signature of binary down to the resolution of $\sim$18 au.
    The TM1+TM2 continuum
    map shows a spiral-like distribution surrounding the central protostar. The continuum visibility profile suggests the presence of three components with distinct size scales, including an extremely compact circumstellar disk with a radius of $\sim 12\rm ~au$. 
    \item Significant chemical stratification is exhibited by the dense gas tracers $\rm C^{18}O$, $\rm N_2D^+$, and $\rm DCO^+$. 
    The spatial distribution of the C$^{18}$O emission peaks toward the continuum position, exhibiting a spatially compact distribution with a radius of  $\sim$240 au, while those of the deuterated molecules show anti-correlation with C$^{18}$O.
    This indicates that the central protostar is extremely young, and has just started heating its surrounding. 
    The bulk of the gas in the core remains colder than the CO evaporation temperature.
    \item {The envelope kinematics  traced by the C$^{18}$O emission exhibits rotational feature with spin-up toward the central source.}  
    {The observed PV diagram is reasonably reproduced by the models with rotating and infalling envelope surrounding a central star with the mass of $\sim${ 0.08}--0.1 $M_{\sun}$.}
    \item The collimated outflow is traced by the CO, SiO, SO, and H$_2$CO emission.
    The axes of the blueshifted and redshifted outflows are misaligned by $\sim$30$^{\circ}$.
    The tips of the lobes exhibit a pair of bow-like patterns, implying that the each lobe consists of two lobes with slightly different position angles.
    The inclination angle of the outflow axes, as determined by the wide-opening angle wind model, is  $ \sim$ {40}$^{\circ}$ from the plane of the sky for both blue- and redshifted lobes. 
    \item The EHV jets with projected velocities of $\gtrsim \rm 20~km~s^{-1}$ have been observed in the $\rm CO$, $\rm SiO$, and $\rm SO$ lines, extending up to $\sim 720\rm ~au$. 
    The axes of the blueshifted ane redshifted jets are also misaligned by $\sim$30$^{\circ}$.
    The inclination angle, mean deprojected jet velocity, and period of velocity variation derived from the shock forming model are $i{\sim}$ {30}$^{\circ}$, $V_j {\sim}$ {63--66} km s$^{-1}$, and $P {\sim}$ {11} yr. 
    \item The jet mass loss rate was estimated to be $\dot{M}_{\rm jet}$$\sim${9.2}${\times}$10${ ^{-8}}$ $M_\odot$ yr$^{-1}$. Comparing with the mass accretion rate derived from the bolometric luminosity, $\dot{M}_{\rm acc}{\sim}${ 7.3}${\times}$10$^{ {-7}}$ $M_{\odot}$ yr$^{-1}$, the magnetic lever arm parameter ${\lambda}_{\phi}{\approx}\dot{M}_{\rm acc}/\dot{M}_{\rm jet}$ becomes $\sim${8.0}. For the central star with a mass of $\sim${0.1} $M_{\odot}$, the footpoint radius of the jet, having a velocity of $\sim${65} km s$^{-1}$, is calculated to be $\sim${0.28} au, which is { within twice of} the dust sublimation radius of the { 1.15} $L_{\odot}$ protostar. This implies that the jet could be launched from the outer edge of the dust sublimation zone.
    \item Given the absence of the binary signature, the two pairs of outflows and the EHV jet are likely to be driven by a single protostar, although an embedded binary with a separation smaller than $\sim$20 au cannot be excluded.
    The varying position angles of the outflows and jets can be attributed to the reorientation of the ejection axes within the turbulent core, where the distribution of angular momentum is nonuniform.  
    \item The origin of the deflection observed in the outflows and jet could be 1) Lorentz forces between the external magnetic field and the current in the outflow lobe or 2) turbulent accretion in moderately magnetized core. 
    The latter scenario is favorable because this scenario can also explain  the multiple pairs of lobes and the jets exhibiting varying orientations. 
\end{enumerate}

\begin{acknowledgments}
M.Y. acknowledges support from the National Science and Technology Council (NSTC) of Taiwan with grant NSTC 112-2124-M-001-014 and NSTC 113-2124-M-001-008. This work was supported by NAOJ ALMA Scientific Research Grant Code 2022-22B. 
We express our gratitude to Drs. Y. Aso and N. Ohashi for their invaluable suggestions regarding the envelope modeling.
This paper makes use of the following ALMA data: ADS/JAO.ALMA\#2017.1.00707.S. ALMA is a partnership of ESO (representing its member states), NSF (USA) and NINS (Japan), together with NRC (Canada), MOST and ASIAA (Taiwan), and KASI (Republic of Korea), in cooperation with the Republic of Chile. The Joint ALMA Observatory is operated by ESO, AUI/NRAO and NAOJ. The National Radio Astronomy Observatory is a facility of the National Science Foundation operated under cooperative agreement by Associated Universities, Inc. 
\end{acknowledgments}

\facility{ALMA, SMA, \emph{Herschel}, \emph{WISE}, JCMT}
\software{CASA \citep{McMullin2007}, CARTA \citep{CARTA2021}, $\tt PRIISM$ \citep{Nakazato2020}, Numpy \citep{numpy2006, numpy2011}, Astropy \citep{astropy2018, astropy2023}, Scipy \citep{scipy2020}, Matplotlib \citep{mpl2007}, Pandas \citep{pandas2010}, Astroquery \citep{astroquery}}

\appendix
\restartappendixnumbering
\section{Molecular Line Channel Maps}
\label{sec:channel_maps}
\begin{figure*}[ht!]
\includegraphics[width=\textwidth]{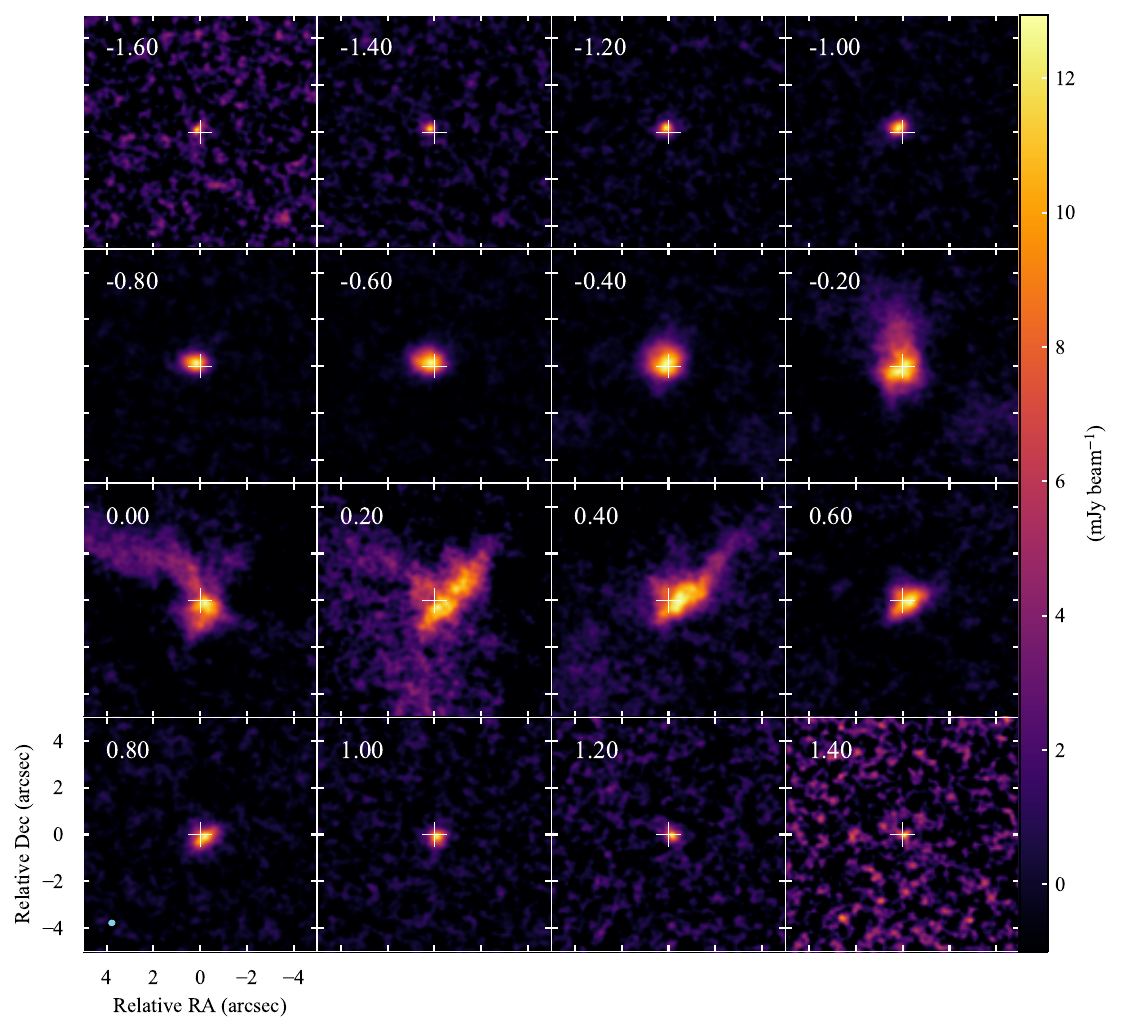}
\caption{Channel maps of C$^{18}$O emission.
Velocity offset from $v_{\rm{sys}}=1.4\rm ~km~s^{-1}$ is given in each panel.
Cross indicates the position of the 1.3 mm continuum peak. 
The blue-filled ellipse in the bottom left corner denotes the synthesized beam size. \label{fig:C18O_channel_map}}
\end{figure*}

\begin{figure*}[ht!]
\includegraphics[width=\textwidth]{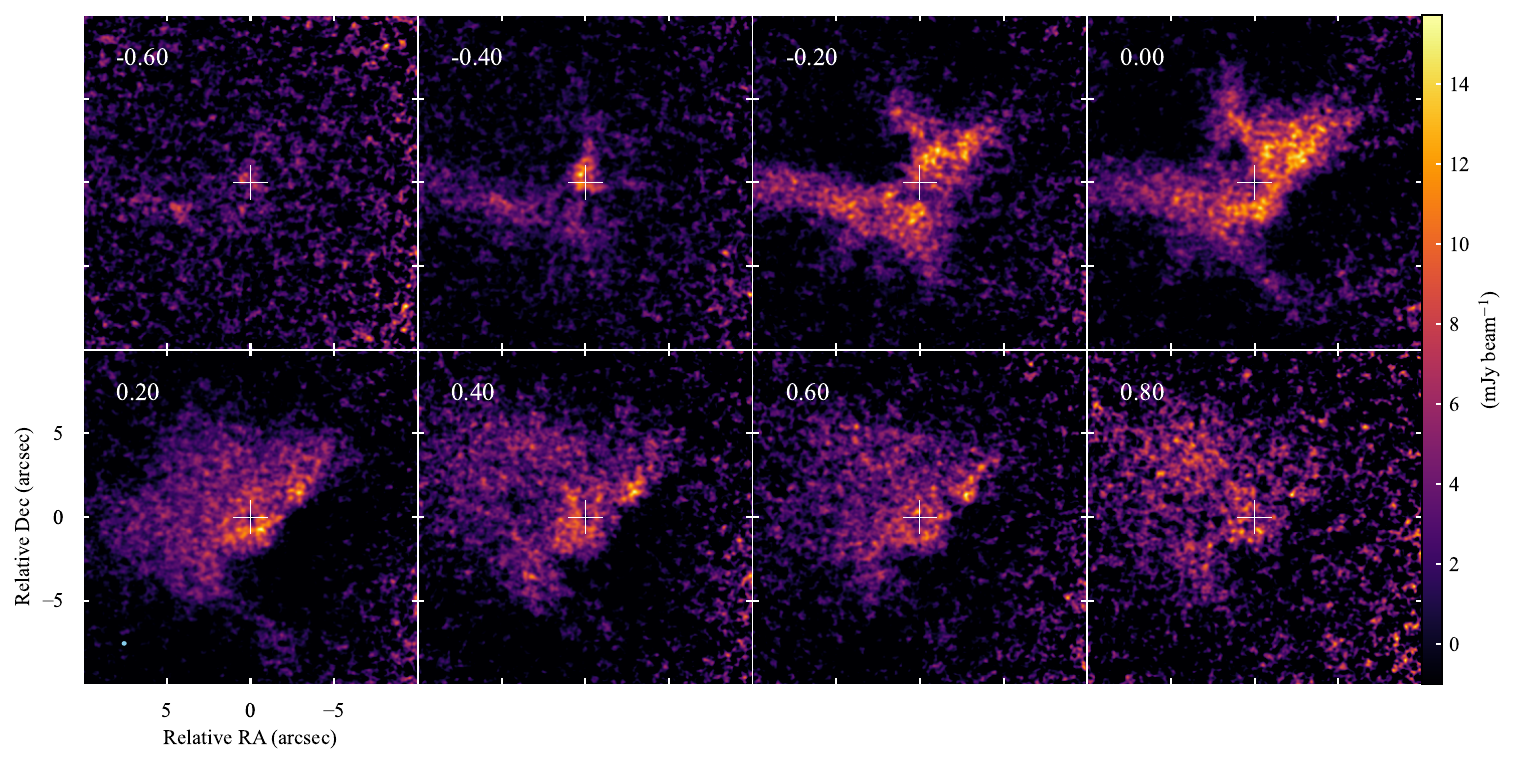}
\caption{Same as Figure \ref{fig:C18O_channel_map} but for $\rm N_2D^+$. \label{fig:N2Dp_channel_map}}
\end{figure*}

\begin{figure*}[ht!]
\includegraphics[width=\textwidth]{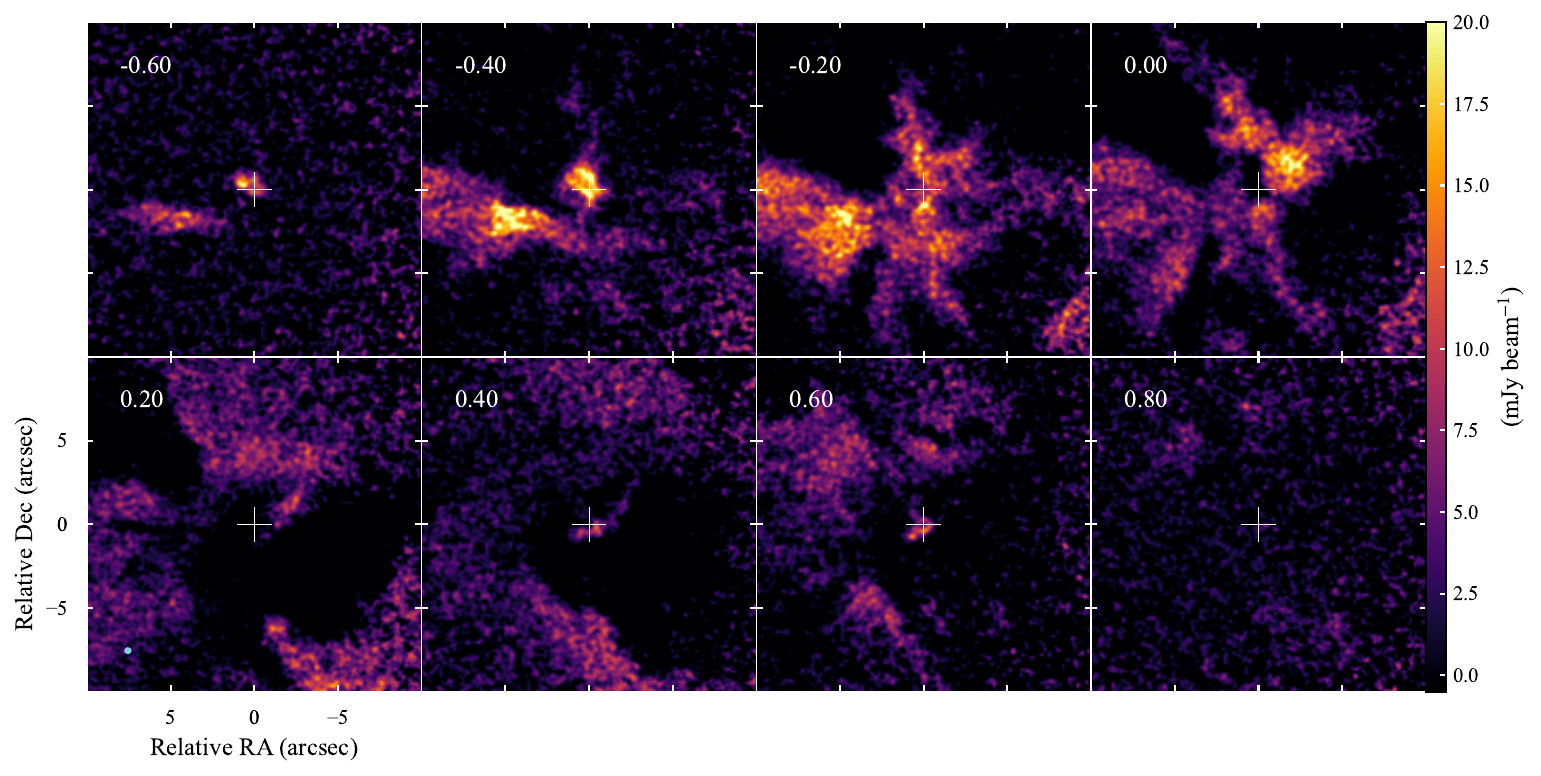}
\caption{Same as Figure \ref{fig:C18O_channel_map} but for $\rm DCO^+$. \label{fig:DCOp_channel_map}}
\end{figure*}

\begin{figure*}[ht!]
\includegraphics[width=\textwidth]{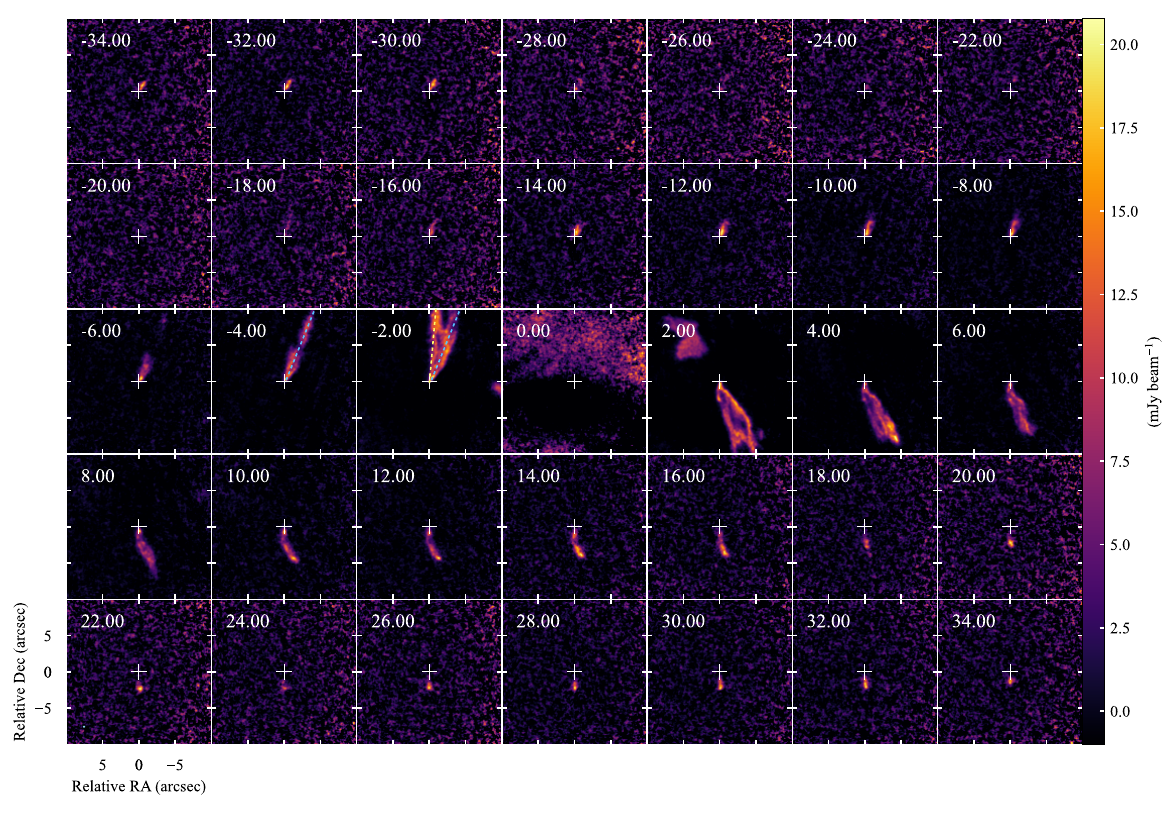}
\caption{Same as Figure \ref{fig:C18O_channel_map} but for $\rm CO$. 
Yellow and cyan dashed lines denote the eastern wall (P.A. $\sim$~$-$5$^{\circ}$) and the western wall (P.A. $\sim$~$-$22$^{\circ}$), respectively, of the northern lobe.\label{fig:CO_channel_map}}
\end{figure*}

\begin{figure*}[ht!]
\includegraphics[width=\textwidth]{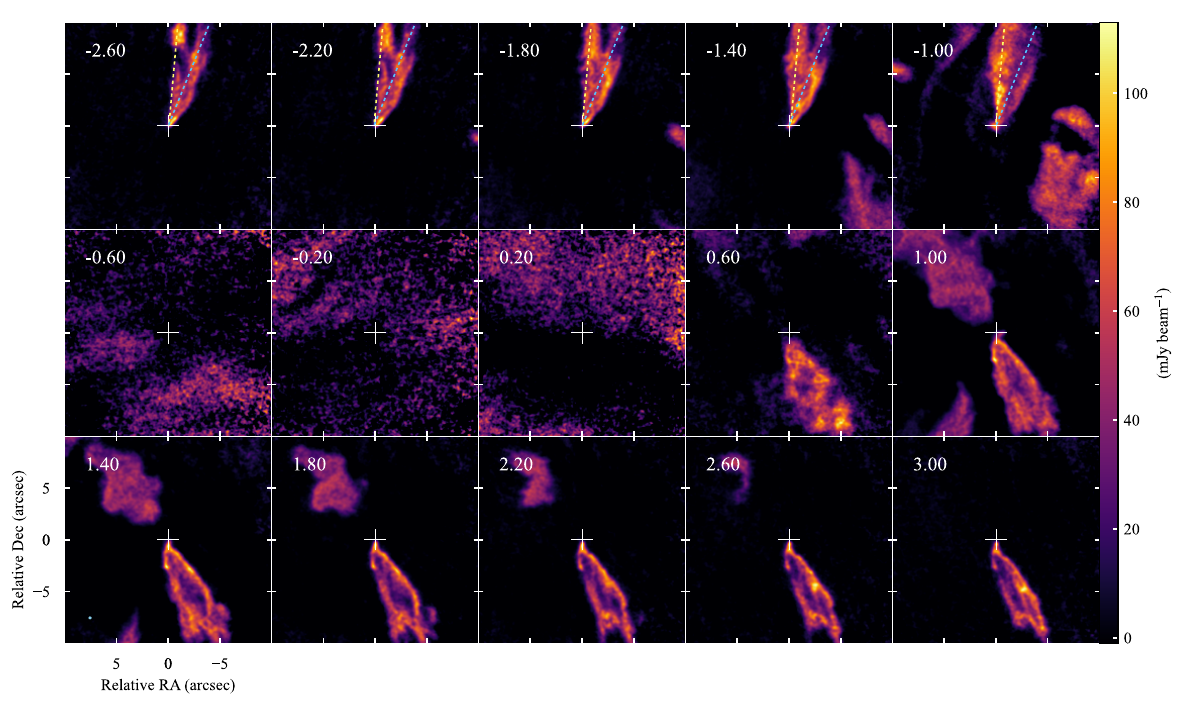}
\caption{Same as Figure \ref{fig:C18O_channel_map} but for the low-velocity channels in the $\rm CO$ line. 
Yellow and cyan dashed lines denote the eastern wall (P.A. $\sim$~$-$5$^{\circ}$) and the western wall (P.A. $\sim$~$-$22$^{\circ}$), respectively, of the northern lobe.
\label{fig:CO_channel_map_lowvel}}
\end{figure*}

\begin{figure*}[ht!]
\includegraphics[width=\textwidth]{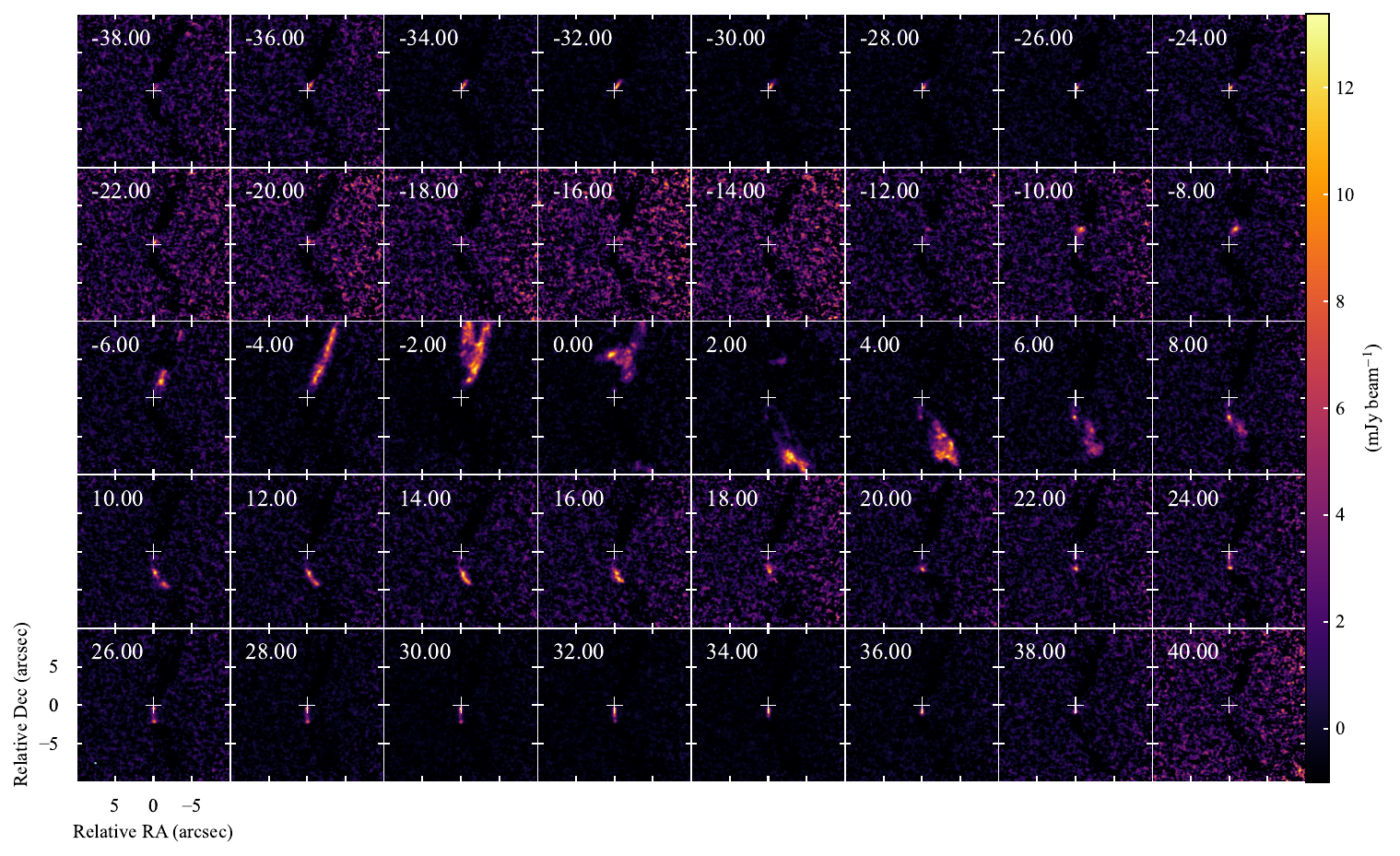}
\caption{Same as Figure \ref{fig:C18O_channel_map} but for $\rm SiO$. \label{fig:SiO_channel_maps}}
\end{figure*}

\begin{figure*}[ht!]
\includegraphics[width=\textwidth]{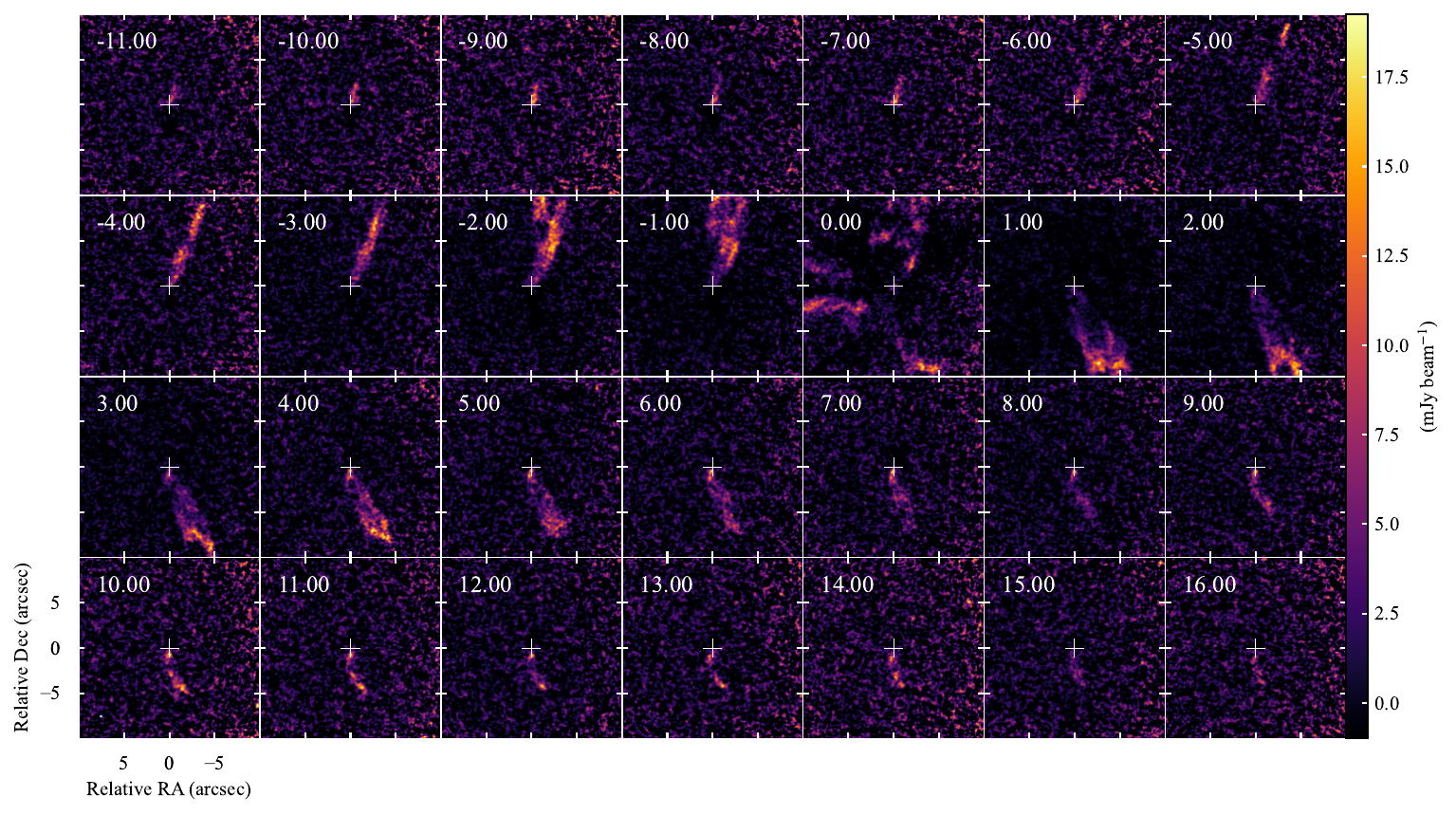}
\caption{Same as Figure \ref{fig:C18O_channel_map} but for $\rm SO$. \label{fig:SO_channel_maps}}
\end{figure*}


\begin{figure*}[ht!]
\includegraphics[width=\textwidth]{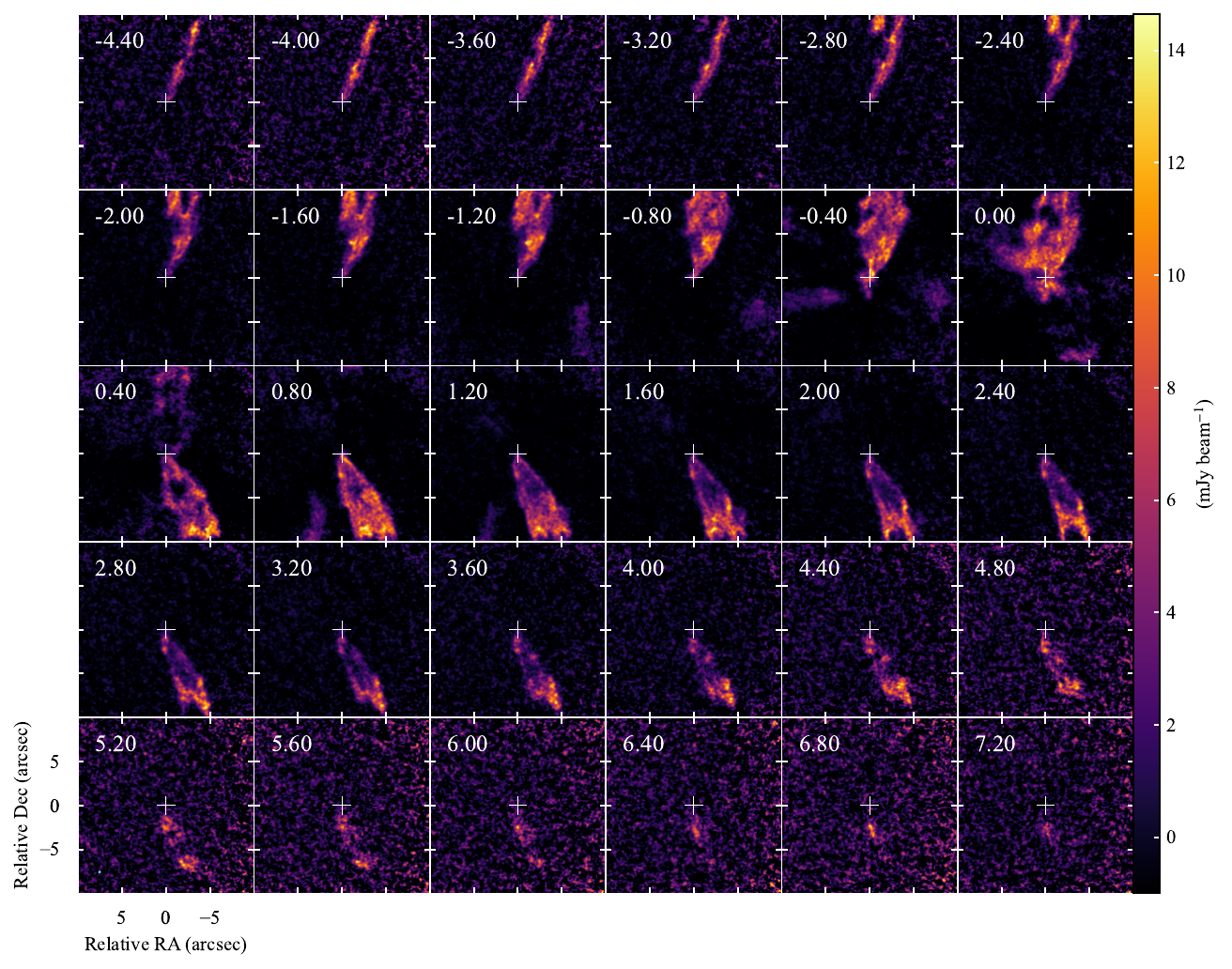}
\caption{Same as Figure \ref{fig:C18O_channel_map} but for $\rm H_2CO~3_{0,3}$--2$_{0,2}$ \label{fig:H2CO303_channel_maps}}
\end{figure*}

\begin{figure*}[ht!]
\includegraphics[width=\textwidth]{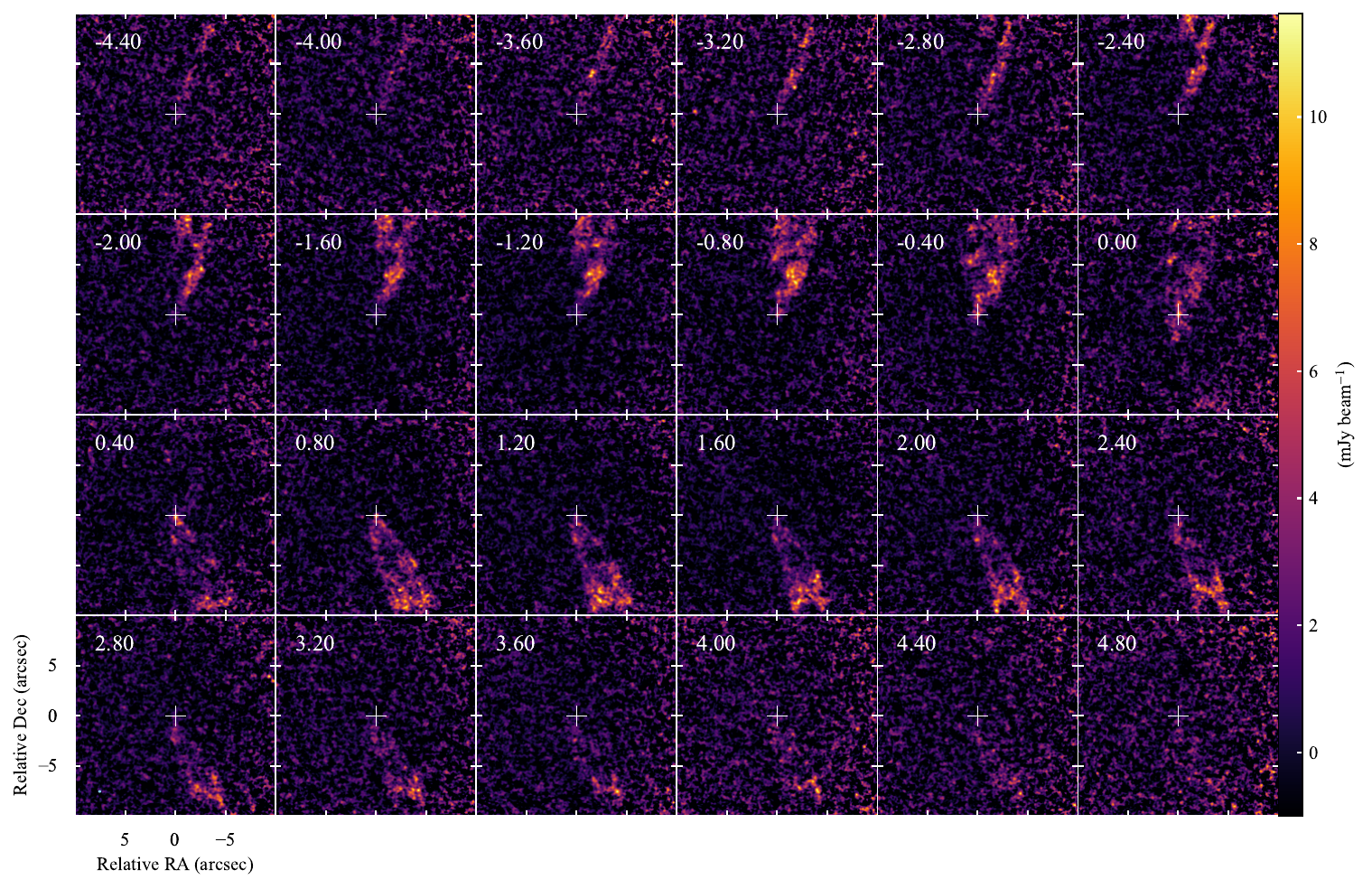}
\caption{Same as Figure \ref{fig:C18O_channel_map} but for $\rm H_2CO~3_{2,2}$--2$_{2,1}$. \label{fig:H2CO322_channel_maps}}
\textbf{}\end{figure*}

The velocity channel maps of the dense gas tracers ($\rm C^{18}O$, $\rm N_2D^+$, and $\rm DCO^+$) are presented in Figures \ref{fig:C18O_channel_map}-\ref{fig:DCOp_channel_map}, while those of the outflow tracers ($\rm CO$, $\rm SO$, $\rm SiO$, $\rm H_2CO~3_{0,3}$ and $\rm H_2CO~3_{2,2}$) are presented in Figures \ref{fig:CO_channel_map}-\ref{fig:H2CO322_channel_maps}.

\section{Derivation of Position-Velocity Model}
\label{sec:pv_derivation}
We describe the calculation procedure of a position-velocity model for a flared envelope with conical cavities and an embedded circumstellar disk at an inclination angle $i$ relative to the edge-on configuration. We assume that the envelope and inner disk are rotating and infalling. Cartesian coordinates within the envelope-disk system are adopted with the x-axis along the projected major axis of the envelope, the y-axis along the line of sight, and the z-axis orthogonal to both the x- and y-axes. The origin is defined as the center of the envelope. 

We define the deprojected coordinates of $(X,\ Y,\ Z)$ as $(X',~Y',~Z')$. The relationship between the coordinates is given by:
\begin{equation}
X' = X,\ \ Y' = Y \cos i - Z \sin i,\ \ Z' = Y \sin i + Z \cos i
\end{equation}
The distances from $(X', Y', Z')$ to the axis of rotation $r_{\rm ax}$ and to the origin $r_{\rm cen}$ can then be expressed in terms of the projected coordinates as:
\begin{equation}
r_{\rm ax} = \sqrt{(X')^{2} + (Y')^{2}},\ \ r_{\rm cen} = \sqrt{(X')^{2}+(Y')^{2}+(Z')^{2}}
\end{equation}

If the angle of cavity $\theta_{cav}$ is very large, the envelope-disk system is almost a spatially-thin disk, specific angular momentum $l$ is roughly conserved, and the centrifugal barrier $R_{0}$ as a function of $Z'$ is almost constant. Hence, the magnitudes of the infall velocity $v_{\rm inf}$ and rotational velocity $v_{\rm rot}$ can be written as:
\begin{equation}
|\vec{v_{\rm inf}}| = \sqrt{\frac{2GM_{\star}}{r_{\rm cen}} - (\frac{l}{r_{\rm ax}})^{2}}
\end{equation}
and 
\begin{equation}
|\vec{v_{\rm rot}}| = \frac{l}{r_{\rm ax}}
\end{equation}
where $G$ and $M_{\star}$ stand for the universal gravitational constant and the mass of the protostar (plus the inner disk), respectively. The centrifugal barrier is thus:
\begin{equation}
r_{\rm CB} = \frac{l^2}{2GM{\star}}
\end{equation}

Within any circular cross-section area at an arbitrary $Z'$, we define the azimuthal angle $\phi$ and polar angle $\theta$ such that:
\begin{equation}
\sin \phi = \frac{Y'}{r_{\rm ax}}, \ \ 
\cos \phi = \frac{X'}{r_{\rm ax}} 
\end{equation}
and 
\begin{equation}
\sin \theta = \frac{Z'}{r_{\rm cen}}, \ \
\cos \theta = \frac{r_{\rm ax}}{r_{\rm cen}}
\end{equation}

The radial velocity $v^{\rm los}$ has two components, the rotational and infall velocities projected to the line of sight. We first consider the component provided by the rotational velocity, which can be given as:
\begin{equation}
\vec{v_{\rm rot}} = |\vec{v_{\rm rot}}| \ (\sin \phi \ \hat{\mathbf{x'}} + \cos \phi \ \hat{\mathbf{y'}})
\end{equation}
where $\hat{\mathbf{x'}} = \langle 1,\ 0,\ 0 \rangle$ and $\hat{\mathbf{y'}} = \langle 0,\ \cos i,\ -\sin i \rangle$ are unit vectors along the $x'$- and $y'$-axes, respectively. Projecting the rotational velocity onto the y-axis gives:
\begin{equation}
v^{\rm los}_{\rm rot} = |\vec{\rm v_{rot}}| \ \cos \phi \cos i
\end{equation}
Next, we consider the infall velocity, which is given by: 
\begin{equation}
\vec{v_{\rm inf}} = |\vec{v_{\rm inf}}|(\cos \theta \cos \phi \ \hat{\mathbf{x'}} 
+ \cos \theta \sin \phi \ \hat{\mathbf{y'}} - \sin \theta \ \hat{\mathbf{z'}}) 
\end{equation}
where $\hat{\mathbf{z'}} = \langle 0,\ \sin i,\ -\cos i \rangle$. Projecting the infall velocity onto the line of sight gives:
\begin{equation}
v^{\rm los}_{\rm inf} = |\vec{v_{\rm inf}}| (\cos \theta \sin \phi \cos i - \sin \theta \sin i)
\end{equation}

The observed intensity $S_v$ is assumed to be proportional to the column density. We suppose that the disk surface density follows a power law with respect to $r_{\rm{cen}}$. In this paper, we assumed an exponent of $p = -1.5$ (i.e., $\Sigma \propto R_{\rm{cen}}^{-1.5}$). Inside the centrifugal barrier, we cap the surface density to $r_{\rm CB}^{-1.5}$ to avoid dividing by zero. If the following conditions due to the geometric boundaries are not satisfied, zero surface density is assigned to that coordinate:
\begin{enumerate}
\item $r_{\rm cen}(X, Y, Z) \leq r_{\rm env}$ and $\theta_{\rm ref}=|\sin^{-1}(|\frac{Z'}{r_{\rm cen}}|)| \leq 90^{\circ} - \frac{\theta_{\rm cav}}{2} $   if $r_{\rm ax} \leq r_{\rm CB}$
\item $|Z'| \leq R_0 \cot{\frac{\theta_{cav}}{2}}$  if $r_{\rm ax} > r_{CB}$
\end{enumerate}
We then calculated the intensities in the simulated three-dimensional (two spatial, one spectral) data cube by integrating along the line of sight with steps of the velocity resolution $\Delta v = 0.1\rm ~km~s^{-1}$. On each simulated channel map, values enclosed by pixels with size $0''.04$ were averaged. Gaussian convolution was performed on each modeled channel map to simulate the $\rm C^{18}O$ beam size (see Table \ref{tab:lines}). From the simulated data cube, the position-velocity diagrams were then extracted along the projected major and minor axes.

\bibliography{sample631}{}
\bibliographystyle{aasjournal}
\end{document}